\documentclass[fleqn,usenatbib]{mnras}
\usepackage{newtxtext,newtxmath}

\usepackage[T1]{fontenc}

\DeclareRobustCommand{\VAN}[3]{#2}
\let\VANthebibliography\thebibliography
\def\thebibliography{\DeclareRobustCommand{\VAN}[3]{##3}\VANthebibliography}

\usepackage{graphicx}	
\usepackage{amsmath}	
\usepackage{gensymb}

\usepackage[usenames]{xcolor}
\usepackage[normalem]{ulem}

\def\reff{r_{\rm eff}}
\def\fbh{f_{\rm BH}}
\def\Vc{V_{\rm c}}
\newcommand{\msun}{{\rm M}_{\odot}}
\def\rh{r_{\rm h}}
\def\rhn{r_{\rm h,0}}
\def\rhf{r_{\rm h,f}}
\def\rj{r_{\rm J}}
\def\rjn{r_{\rm J,0}}
\def\Mi{M_{\rm i}}
\def\Mbh{M_{\rm BH}}
\def\fesc{f_{\rm e}}
\def\mdotref{\dot{M}_{\rm ref}}
\def\tev{t_{\rm ev}}

\def\pc{{\rm pc}}
\def\myr{{\rm Myr}}

\defcitealias{2003MNRAS.340..227B}{BM03} 
\defcitealias{2023MNRAS.522.5340G}{GG23} 
\defcitealias{2021NatAs...5..957G}{G21} 

\title[BH streams]{Stellar streams from black hole-rich star clusters}

\author[]{
Daniel Roberts$^{1,2,3}$\thanks{E-mail: d.m.roberts@soton.ac.uk (DR); mgieles@icc.ub.edu (MG); d.erkal@surrey.ac.uk (DE).},
Mark Gieles$^{3,4}$, 
Denis Erkal$^{2}$ and 
Jason L. Sanders$^{5}$
\\
$^{1}$School of Physics and Astronomy, University of Southampton, Highfield, SO17 1BJ, UK\\
$^{2}$Department of Physics, University of Surrey, Guildford GU2 7XH, UK\\
$^{3}$Institut de Ci\`{e}ncies del Cosmos (ICCUB), Universitat de Barcelona (IEEC-UB), Mart\'{i} Franqu\`{e}s 1, E08028 Barcelona, Spain\\
$^{4}$ICREA, Pg. Lu\'{i}s Companys 23, E08010 Barcelona, Spain\\
$^5$Department of Physics and Astronomy, University College London, London WC1E 6BT, UK
}

\date{Accepted XXX. Received YYY; in original form ZZZ}

\pubyear{2024}

\begin{document}
\label{firstpage}
\pagerange{\pageref{firstpage}--\pageref{lastpage}}
\maketitle

\begin{abstract}
Nearly a hundred progenitor-less, thin stellar streams have been discovered in the Milky Way, thanks to {\it Gaia} and related surveys. Most streams are believed to have formed from star clusters and it was recently proposed that extended star clusters -- rich in stellar-mass black holes (BHs) -- are efficient in creating streams. To better understand the nature of stream progenitors, we quantify the differences between streams originating from star clusters with and without BHs using direct $N$-body models and a new model for the density profiles of streams based on time-dependent escape rates from clusters: the Quantifying Stream Growth (QSG) model. QSG facilitates the rapid exploration of parameter space and provides an analytic framework to understand the impact of different star cluster properties and escape conditions on the structure of streams. Using these models it is found that, compared to streams from BH-free clusters on the same orbit, streams of BH-rich clusters: (1)  are approximately five times more massive; (2) have a peak density three times closer to the cluster $1~{\rm Gyr}$ post-evaporation (for orbits of Galactocentric radius $\gtrsim 10~{\rm kpc}$), and (3) have narrower peaks and more extended wings in their density profiles. We discuss other observable stream properties that are affected by the presence of BHs in their progenitor cluster, namely the width of the stream, its radial offset from the orbit, and the properties of the gap at the progenitor's location. Our results provide a step towards using stellar streams to constrain the BH content of evaporated (globular) star clusters. 
\end{abstract}


\section{Introduction}\label{sec:intro}
Stellar streams are the debris of evaporated star clusters \citep[for example,][]{2001ApJ...548L.165O, GD1_disc} and accreted dwarf galaxies \citep*[for example,][]{1994Natur.370..194I} and are found in both the inner \citep*{2019ApJ...872..152I} and outer halo \citep[for example,][]{Belokurov+2006,Newberg+2010,2018ApJ...862..114S} of the Milky Way (MW) as well as in other galaxies \citep[for example,][]{2001Natur.412...49I,2010AJ....140..962M,2023A&A...671A.141M}.
In recent years, there has been a significant uptick in the discovery rate of streams in the  MW halo \citep[see the review by][]{2025NewAR.10001713B}, thanks to the advent of the ESA {\it Gaia} space telescope \citep*[for example,][]{2018MNRAS.481.3442M, 2019ApJ...872..152I} and deep, wide-area photometric surveys \citep[for example,][]{Koposov+2014,Bernard+2016,2018ApJ...862..114S}. Streams are powerful tools in studies of the MW: their shapes provide important constraints on the  gravitational potential of the MW (for example, \citealt{1995MNRAS.275..429L}; \citealt*{2010ApJ...712..260K}; \citealt{2015ApJ...803...80K, Bovy+2016}; \citealt{Erkal+2019}; \citealt{2023MNRAS.521.4936K}) and their chemistry and orbits help to reconstruct the  assembly history of the MW \citep[for example,][]{Bonaca+2021,Li+2022}. 

The precise astrometric data provided by {\it Gaia} has tightly constrained the orbits of the observed GCs and stellar streams \citep[for example,][]{Li+2022,2025NewAR.10001713B}.
For streams originating from star clusters, the orbit combined with the Galactic potential provides  constraints on the mass loss history of the  progenitor cluster (\citealt{2003MNRAS.340..227B}, hereafter \citetalias{2003MNRAS.340..227B}; \citealt{2023MNRAS.522.5340G}, hereafter \citetalias{2023MNRAS.522.5340G}; \citealt*{Chen+2024}). 

The narrow width and low velocity dispersion of the GD-1 stream \citep{2010ApJ...712..260K} and the chemistry of its stars \citep*{2022MNRAS.515.5802B} argue for a star cluster origin. However, \citet*{2020MNRAS.494.5315D} noted that the initial stellar mass in the GD-1 stream is about five times larger than the estimated maximum mass of a star cluster that can evaporate on that orbit, based on $N$-body calculations of Roche-filling star clusters evaporating in a Galactic tidal field (\citetalias{2003MNRAS.340..227B}). Curiously, some globular clusters (GCs) with much closer pericentric passages than the GD-1 stream have no noticeable tidal tails associated with them \citep*{2018MNRAS.473.2881K}. This suggests that, in addition to the time spent on that orbit, an orbit-independent parameter is required to explain the variation in mass-loss rates of GCs.

\citet[][hereafter \citetalias{2021NatAs...5..957G}]{2021NatAs...5..957G} showed that the additional parameter is most likely the dynamical effect of stellar-mass black holes (BHs) in the progenitor star cluster. $N$-body models of tidally limited star clusters with different initial masses \citep{Pavlik2018BHretention} and densities (\citetalias{2021NatAs...5..957G}; \citealt{2024MNRAS.527.7495W}) retain a different fraction of BHs and therefore evolve to have different BH populations today \citep{10.1093/mnras/stt628}, impacting the mass-loss of the cluster \citep{2011ApJ...741L..12B,2019MNRAS.487.2412G}. \citetalias{2021NatAs...5..957G} build upon this to show that the stream associated with the halo GC Palomar 5 (hereafter Pal 5) -- just like the GD-1 stream -- also contains more mass than can be explained by the models of the evaporation of star clusters without BHs 
(\citetalias{2003MNRAS.340..227B}). Because this is the most prominent stream with a known progenitor, \citetalias{2021NatAs...5..957G} attempted to reproduce the observed properties of both the cluster and the stream with $N$-body simulations. They found that both the peculiar, large half-light radius of $\reff\simeq20\,\pc$ of Pal 5, as well as the mass in the stream can only be reproduced if the cluster contains a BH population, constituting  a  fraction of $\fbh\simeq0.2$ of the total present-day cluster mass. From these models, \citetalias{2021NatAs...5..957G} found that both the mass in the tails as well as $\reff$ correlate with  $\fbh$ and concluded that BH-rich GCs are the likely progenitors of cold streams. The higher mass-loss rate of GCs with BHs also helps to explain the shape of the GC mass function and the distribution of nitrogen-rich stars in the inner halo that are believed to originate from GCs (\citetalias{2023MNRAS.522.5340G}).

\begin{figure}
    \centering
    \includegraphics[width=.45\textwidth]{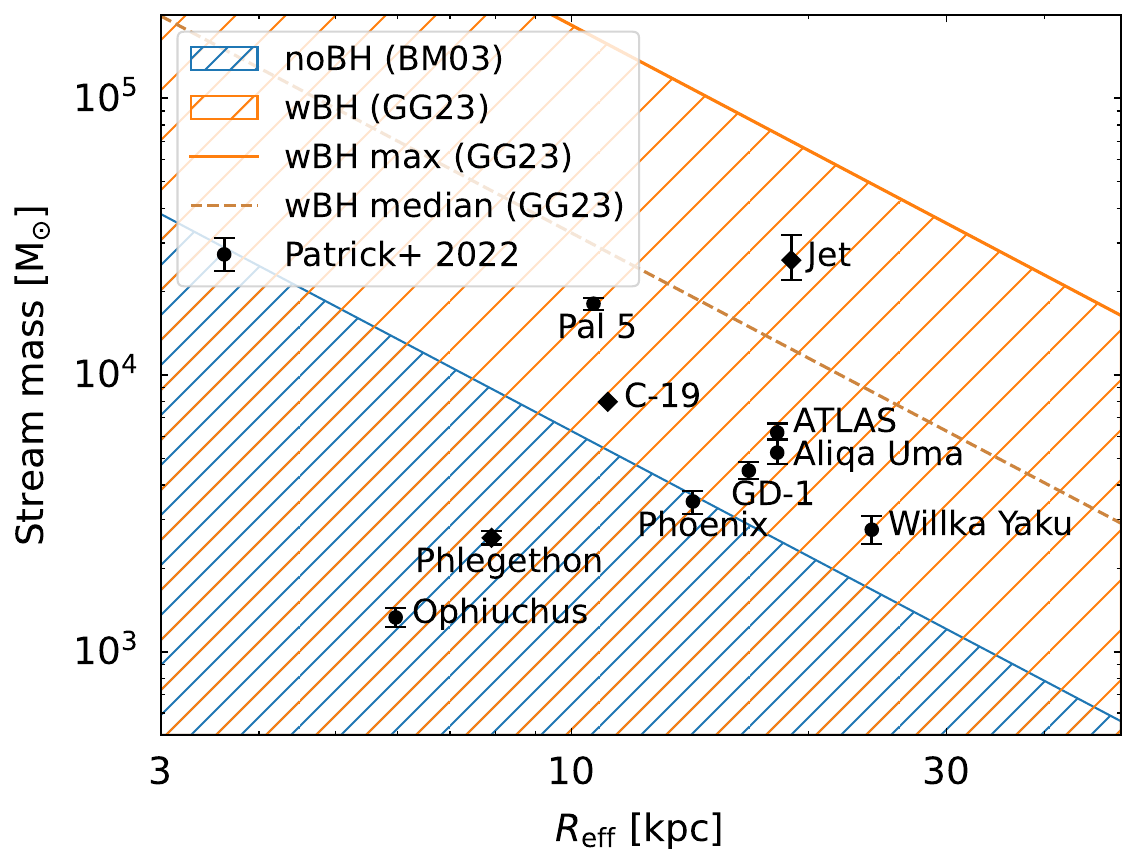}
    \caption{The mass of the streams from \citet{2022MNRAS.514.1757P}, that are believed to have had GC progenitors which have now evaporated, as a function of the Galactocentric radius of the equivalent circular orbit with the same average mass-loss rate, $R_{\rm eff}$ (\citetalias{2003MNRAS.340..227B}). In addition, we include Jet \citep{2022AJ....163...18F}, C-19 \citep{2022Natur.601...45M}, and Phlegethon \citep{Ibata_2018} (denoted by the diamond markers) which are also believed to have GC progenitors. The blue hatched region denotes the area of parameter space that can be populated by GCs without BHs that can evaporate within $10~{\rm Gyr}$. The orange hatched region denotes the area of parameter space that can be populated by GCs with BHs that can evaporate within $10~{\rm Gyr}$ and the dashed (solid) orange line marks the median (maximum) mass of a GC with BHs and an evaporation time of $10~{\rm Gyr}$ (see text for details). 
    The error bars do not take into account the possibility of unobserved low density extensions to the streams and therefore these data points should be regarded as lower bounds on the progenitor mass.
    }
    \label{fig:Mass Patrick Comp}
\end{figure}

In addition to GD-1 and the Pal 5 stream, there are more streams with masses above the maximum masses of clusters without BHs that can evaporate on their orbits. Figure \ref{fig:Mass Patrick Comp} displays the masses of the streams included in \citet*{2022MNRAS.514.1757P} that are believed to have had GC progenitors which have now evaporated as a function of the Galactocentric radius of their equivalent circular orbits with the same average mass-loss rate ($R_{\rm eff} = R_{\rm p} (1+\varepsilon)$, where $R_{\rm p}$ is the Galactocentric radius at pericentre and $\varepsilon$ is the orbital eccentricity, \citetalias{2003MNRAS.340..227B}). In addition, we include Jet, C-19, and Phlegethon streams which are believed to have GC progenitors. The orbital parameters used to calculate $R_{\rm eff}$ for the streams in \citet{2022MNRAS.514.1757P} were obtained from \citet{Li+2022} for all streams except GD-1 and Pal 5 which used the values from \citet{Bonaca_2020} and \citet{2015ApJ...803...80K}, respectively. The mass estimates and orbital parameters for Jet, C-19, and Phlegethon were taken from \citet{2022AJ....163...18F}, \citet{2022Natur.601...45M}, and \citet{Ibata_2018}, respectively. The lines show model predictions (\citetalias{2023MNRAS.522.5340G}) for the initial mass, after stellar evolution, of GCs with an evaporation time of $10~{\rm Gyr}$ without BHs (blue) and with BHs (orange). In Appendix~\ref{app:mt} we provide details on how these limiting masses were derived from the GG23 model. The `wBH' lines correspond to clusters that have BHs during their entire evolution, and `noBH' lines are for models that either quickly ejected BHs early in the evolution because of a short initial relaxation time (low GC mass/high density), or never had BHs.

Apart from Ophiuchus and Phlegethon, which lie well below the noBH line, and Phoenix which lies just below the noBH line, every other stream's mass exceeds the noBH limit. \citet{2022MNRAS.514.1757P} calculate the mass of the stream by generating a new simulated stellar population from the fitted colour-magnitude diagram, allowing the mass estimate to account for unobserved low mass stars. Yet, these masses are likely lower limits because \citet{2022MNRAS.514.1757P} did not include features offset from the stream track (such as the spur of GD-1), nor corrections for stars outside of their defined ends of the streams which are difficult to pick out from the background. This implies that their mass estimates are lower limits of the initial stellar mass of the streams' progenitor GCs. A particularly stark example is that of GD-1, for which \citet{2022MNRAS.514.1757P} determine the stream mass to be $\sim5\times10^3\,\msun$, whereas \citet{2020MNRAS.494.5315D}  estimate a total mass of $\sim10^4~\msun$ (after stellar evolution mass loss). In addition, it is important to note that the mass estimate of C-19 in \citet{2022Natur.601...45M} is a lower limit as it is expected that the stream extends beyond the observed range. This comparison confirms that most streams evaporated faster than what is expected from models of GCs without BHs, calling for stream formation models that include the effect of BHs.  

Most stream modelling efforts to date have focused on the shape and features such as epicyclic overdensities \citep[for example,][]{2014MNRAS.443..423S, 2014ApJ...795...95B,Fardal+2015,2015ApJ...803...80K} and could therefore adopt a constant escape rate. However, it is well understood that the mass-loss rate is not constant, instead the magnitude of the mass-loss rate decreases as the progenitor evaporates for noBH GCs (\citealt{2000MNRAS.318..753F,2001MNRAS.325.1323B}; \citetalias{2003MNRAS.340..227B}; \citealt{2010MNRAS.409..305L}), while it increases for wBH GCs (\citealt{2011ApJ...741L..12B,2019MNRAS.487.2412G}; \citetalias{2021NatAs...5..957G}), and there is yet to be a study of the dependence of a stream's morphology on the progenitor's mass-loss rate. Motivated by this, the suggestions that BH-rich clusters are the progenitors of (most of the) cold stellar streams (\citetalias{2021NatAs...5..957G}) and that wBH streams should exhibit a gap at the progenitor's position post-evaporation \citep{2020MNRAS.494.5315D}, and also by the availability of more luminosity-based mass estimates and corresponding density profiles of streams (for example, \citealt{2020MNRAS.494.5315D}; \citealt{2022MNRAS.514.1757P}), we here present a model for streams based on a time-dependent mass-loss history of their progenitor clusters. To parameterise this new model, we use direct $N$-body simulations of star clusters evaporating in a Galactic tidal field with and without BHs, to shed light on the nature of stream progenitors by investigating whether the structure of a stream can be used to discriminate between BH-rich and BH-free progenitors and identify features which display a dependence on the retained BH population.

This paper is organised as follows: in Section~\ref{sec:model} we introduce the $N$-body simulations and the model for the density profiles of streams with time-dependent mass-loss histories. In Section~\ref{sec:impact} we discuss the impact of a BH population in the progenitor cluster on stream properties and the discussion and conclusions are presented in Section~\ref{sec:discussion} and Section~\ref{sec:conclusions}, respectively.

\section{A model for streams from time-dependent mass-loss rates}
\label{sec:model}
In this section we present a model for streams forming from clusters on circular orbits with time-dependent mass-loss rates. We first present two $N$-body models of clusters with and without BHs in Section~\ref{ssec:nbody} and then describe the (semi-)analytic model for the stream density profile in Sections~\ref{ssec:mdot}--\ref{ssec:pps}. In Section~\ref{ssec:comp} we compare the stream model to the $N$-body simulations. 

\subsection{$N$-body simulations}
\label{ssec:nbody}
To quantify the effect of a BH population on the resulting stream, we run  $N$-body models of two clusters on the same orbit, where one model contains BHs (`wBH-Nbody') and the other cluster does not (`noBH-Nbody'), the key parameters of these models are summarised in Table \ref{tab:NbodyICs}. We run both simulations with {\sc PeTar}\footnote{\href{https://github.com/lwang-astro/PeTar}{https://github.com/lwang-astro/PeTar}} \citep{wang2020petar}, which includes the effect of stellar and binary evolution \citep{hurley2000SSE, hurley2002BSE} with the recent updates for massive star winds and BH masses from \citet{banerjee2020bse}. We adopt the rapid supernova mechanism by \citet{Fryer_2012}, for which $60$ per cent ($70$ per cent) by number (mass) of the BHs do not receive a natal kick due to fall back, for the adopted stellar initial mass function (IMF) \citep[][in the range $0.1-100\,\msun$]{kroupa2001IMF} and metallicity ($Z =  10^{-3}$, that is, ${\rm [Fe/H]} \simeq -1.1$).  
For the noBH-Nbody model we prevent the formation of BHs by truncating the IMF at $20~\msun$. We adopt a `GD-1 like' orbit: a circular orbit at a Galactocentric radius of $R=20\,$kpc in a singular isothermal sphere (SIS) using the {\sc Galpy} library \citep{bovy2015galpy}\footnote{We use {\sc Galpy}'s pseudo-isothermal sphere with circular velocity of $\Vc = 220\,$km/s at large radii and a core radius of $1\,$pc.}.

\begin{figure}
    \centering
\includegraphics[width=.45\textwidth]{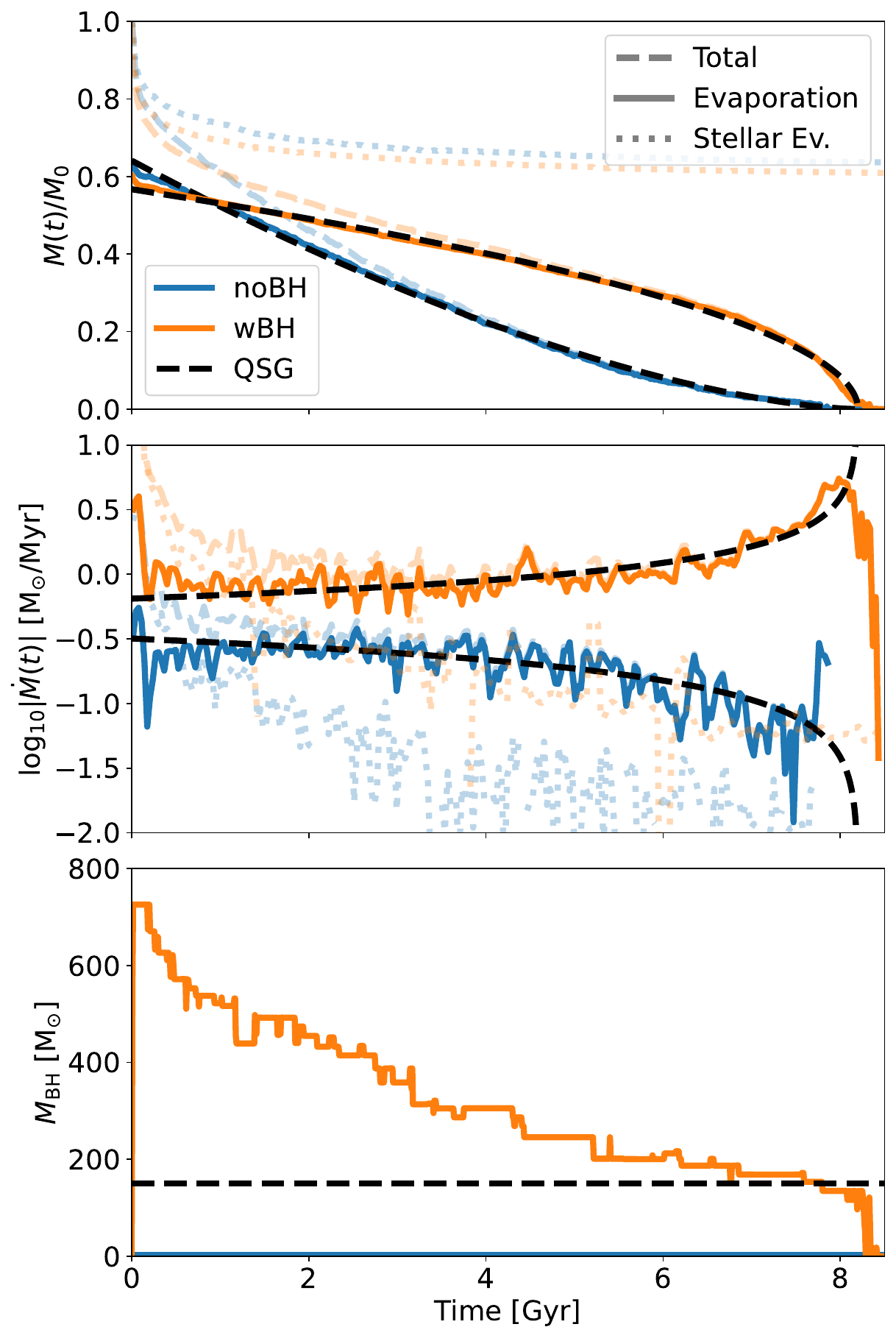}
    \caption{The  evolution of the total cluster mass (\emph{top}), its mass-loss rate (\emph{middle}), and the mass of the black hole population (\emph{bottom}) of the two $N$-body models discussed in Section~\ref{ssec:nbody} (noBH-Nbody 
 in blue and wBH-Nbody in orange), with the contributions from stellar mass loss and evaporation in the tidal field shown separately. The over-plotted black dashed lines are $M(t)$ (equation~\ref{eq:M(t)}) and $\dot{M}(t)$  (equation~\ref{eq:Mdot}), with $\tev=8.2\,$Gyr and $\eta = 0.36; \Mi = 1.66\times10^3\,\msun$ (noBH) and $\eta = -0.94; \Mi = 1.03\times10^4\,\msun$ (wBH). In the bottom panel the over-plotted dashed line corresponds to $M_{\rm BH} = 150~\msun$, the total mass of the retained BH population used in the  wBH-QSG models.}
    \label{fig:Mdot Mt}
\end{figure}

The initial positions and velocities of the stars are drawn from a Plummer model \citep{Plummer_1911} truncated at 20 times the half-mass radius ($\rh$).
We define the initial $\rh$, $\rhn$, in units of the half-mass radius of a Roche-filling cluster ($\rhf$), for which we adopt the value from \citet{1961AnAp...24..369H} of $\rhf=0.15\rj$, where $\rj$ is the Jacobi radius. For the SIS, $\rjn=\left[GM_0/(2\Omega^2)\right]^{1/3}$, where $M_0$ is the initial mass of the cluster and $\Omega=\Vc/R$ is the angular frequency of the orbit. Because clusters expand as a result of stellar mass loss, we start with $\rhn<\rhf$. For the noBH-Nbody cluster we adopt $\rhn=0.7\rhf$ and for the wBH-Nbody cluster we adopt a slightly smaller radius of $\rhn=0.6\rhf$, because this cluster expands more due to the dynamical effect of the BHs following stellar mass loss. What remains to be decided is the initial number of stars ($N$) of both models. We find values for $N$ by iteration, such that both models evaporate approximately at an age of 8\,Gyr, where we define the evaporation time ($\tev$) as the time at which the cluster reaches $0.5\%$ of the initial cluster mass. Our initial estimates are guided by the analytic expressions for $\dot{M}$ and $M(t)$ of \citetalias{2023MNRAS.522.5340G} for clusters with different BH contents. After a few iterations of $N$-body models with different $N$ we settled on $N=4600$ for the noBH-Nbody model and $N=28500$ for the wBH-Nbody model, where the difference is due to the higher mass loss rate of a GC with BHs. The evolution of the total cluster mass, the mass-loss rate and the mass of the BH population of both models is shown in Fig.~\ref{fig:Mdot Mt}. In the next subsection we discuss the resulting streams as well as a generative model for the stream that we base on the mass evolution of these $N$-body models. 

\begin{table}
    \centering
    \begin{tabular}{||c|c|c||}
        \hline
        Value at $t=0$ & noBH-Nbody & wBH-Nbody \\
        \hline\hline 
        $R$ [kpc] & 20 & 20 \\
        $V_{\rm c}$ [km/s] & 220 & 220 \\
        $M_0$ [$\msun$] & 2596 & 18175 \\
        $N$ & 4600 & 28500 \\
        $r_{\rm h,0}$ [${\rm pc}$] & 3.49 & 5.96 \\
        $\tev$ [Gyr] & 7.79 & 8.36 \\
        \hline
    \end{tabular}
    \caption{The initial conditions and the evaporation time of the $N$-body simulations used in this work.}
    \label{tab:NbodyICs}
\end{table}

\subsection{Mass-loss rate}
\label{ssec:mdot}
In this section we develop a semi-analytic model for the streams from clusters with and without BHs and benchmark the results against the $N$-body models from Section~\ref{ssec:nbody}.

We only model the mass-loss due to evaporation, that is, we do not include mass-loss by stellar evolution, which dominates in the early evolution of the GCs. As can be seen in Fig. \ref{fig:Mdot Mt}, stellar evolution dominates the mass-loss of the $N$-body models until $\sim1\,$Gyr. As demonstrated by GG23, the mass-loss rate due to evaporation is well described by a power-law dependence on $M$ of the form 
\begin{equation}\label{eq:mdot powerlaw}
    \dot{M}=-AM^{\eta} = -\frac{\Mi}{(1-\eta)\tev}\left(\frac{M}{\Mi}\right)^{\eta} ,
\end{equation}
where $M$ is the cluster mass, $\Mi<M_0$ is the remaining mass after most stellar evolution mass loss has occurred and $\tev$ is the evaporation time, defined here as the time it takes for the mass to reach $0.005 M_{0}$. From integrating equation~(\ref{eq:mdot powerlaw})  we  obtain the GC mass evolution in time
\begin{equation}\label{eq:M(t)}
    M(t) = \Mi \left(1-\frac{t}{\tev}\right)^{1/(1-\eta)}.
\end{equation}
By substituting this expression for $M(t)$ into equation~(\ref{eq:mdot powerlaw}), or by taking its derivative with respect to time, we find an expression for $\dot{M}(t)$
\begin{equation}\label{eq:Mdot}
    \dot{M}(t) =-\frac{\Mi}{(1-\eta)\tev}\left(1-\frac{t}{\tev}\right)^{{\eta}/({1-\eta})} .
\end{equation}
The mass dependence of $\dot{M}$ is encapsulated in the parameter $\eta$, where the mass evolution of noBH clusters is well described by $\eta \simeq 1/3$ (\citetalias{2003MNRAS.340..227B}) and for wBH clusters, $-1\lesssim\eta\lesssim-1/3$ (\citetalias{2023MNRAS.522.5340G}). The above expressions are  simplified versions of $M(t)$ and $\dot{M}(t)$ expressions recently presented in \citetalias{2023MNRAS.522.5340G}\footnote{They define $\dot{M} \propto M^{1-y}$, such that our $\eta$ relates to their $y$ as  $\eta=1-y$.}. These authors showed with $N$-body models that clusters with lower initial densities retain more of their BHs, and have a smaller (that is,  more negative) $\eta$. For $\eta<0$, the (absolute) mass loss rate increases as the progenitor evaporates, which is the result of the increasing fraction of mass in BHs. \citetalias{2023MNRAS.522.5340G} also show that the constant of proportionality $A$ (which is inversely proportional to $\tev$) in equation~(\ref{eq:mdot powerlaw}) depends on $\eta$, $M_0$ and the strength of the tidal field.

The dependence of $\eta$ on the mass of the BH population, $M_{\rm BH}$, is because there exists a critical $\fbh$ for tidally limited GCs (few percent) at which the stellar mass-loss rate equals the BH mass-loss rate and therefore $\fbh$ remains constant \citep{10.1093/mnras/stt628}. If $\fbh > f_{\rm BH,\, crit}$ then $\fbh$ increases which leads to an accelerating mass-loss rate ($\eta<0$) and a BH-dominated cluster (\citealt{2011ApJ...741L..12B,2019MNRAS.487.2412G}; \citetalias{2021NatAs...5..957G}). On the other hand, if $\fbh < f_{\rm BH,\, crit}$ then $\fbh$ decreases, leading to a decelerating mass-loss rate ($\eta>0$, \citealt{2000MNRAS.318..753F,2001MNRAS.325.1323B}; \citetalias{2003MNRAS.340..227B}; \citealt{2010MNRAS.409..305L}). It is important to note that a  GC without BHs with a low initial density can have a similar high mass-loss rate as a wBH GC with the same mass, but \citetalias{2021NatAs...5..957G} show that this area of parameter space is extremely small and as such we consider an accelerating mass-loss rate to be the result of a retained stellar-mass BH population.

Rather than using the full expressions from \citetalias{2023MNRAS.522.5340G}, we here stick to the simpler expressions from above and find the values of $\Mi$ and $\eta$ that are needed to describe the two $N$-body models. The mass and mass-loss history for the $N$-body models and the analytic approximations are displayed in the top two panels of Fig. \ref{fig:Mdot Mt}. $\Mi$ and $\eta$ are determined by a least-squares fit to the mass evolution ($M(t)$) and the mass-loss rate ($\dot{M}(t)$), excluding the contribution by stellar evolution. We set an upper bound for $\Mi$ of the total stellar mass in the $N$-body model post-evaporation. For the noBH model we then find  $\Mi\simeq0.64 M_0\simeq 1.66\times 10^3~\msun$ and $\eta\simeq0.36$. A similar mass dependence of the mass-loss rate was found previously in models of clusters without BHs (\citetalias{2003MNRAS.340..227B}; \citealt*{2010MNRAS.409..305L}). For the wBH model we find  $\Mi\simeq0.56 M_0\simeq1.03\times10^4\,\msun$\footnote{Despite the fact that the IMF was truncated at different upper masses ($100\,\msun$ for wBH and $20\,\msun$ for noBH), the ratio $\Mi/M_0$ is similar in both cases because the fraction of the initial mass above $20\,\msun$ that ends up in BHs is $\sim0.45$, that is, only slightly lower than the remaining mass fraction of the IMF below $20\,\msun$.} and $\eta\simeq-0.94$. This negative $\eta$ causes mass loss to accelerate as the progenitor evaporates, as found here (see orange lines in top and middle panels of Fig.~\ref{fig:Mdot Mt}) and also in other models of evaporating clusters with BHs (\citealt{2019MNRAS.487.2412G, 2020MNRAS.491.2413W}; \citetalias{2023MNRAS.522.5340G}).

We plot the resulting analytic expressions for $M(t)$ and $\dot{M}(t)$ in Fig.~\ref{fig:Mdot Mt}. These expressions ensure that a GC will evaporate at a chosen $t_{\rm ev}$ (informed by $N$-body simulations), which is key when examining the growth of the gap that forms at the progenitor's position post-evaporation.

\subsection{The Quantifying Stream Growth (QSG) model}\label{ssec:stream growth}

To investigate the differences in streams resulting from the noBH and wBH clusters, a model of the growth of streams is required. Here we introduce a new model that follows the formalism of \citet{erkal2015threephase}. It adopts a reference frame that is centred on the cluster and co-rotates with the orbit, where the $x$-axis points towards the galactic anti-centre, the $y$-axis points along the orbit, and the $z$-axis is along the angular momentum vector of the orbit, perpendicular to the orbital plane. We restrict ourselves to a cluster on a circular orbit, with galactocentric radius $R$ and circular velocity $V_{\rm c}$, within a spherical potential. Stars are then assumed to escape through the Lagrange points, offset from the centre of the cluster along $x$ by a distance $\pm \fesc \, r_{\rm J}(M(t))$, where $\fesc$ is a dimensionless constant of order unity to be determined, and they are released with some initial velocity offset $\Vec{\Delta v} = (\Delta v_{x}, \Delta v_{y}, \Delta v_{z})$. In this model the velocity offset in the $y$-direction ($\Delta v_y$) is related to the escape radius by $\epsilon$, a dimensionless free parameter of order unity, such that $\Delta v_y$ is the random component of the velocity (in the galactocentric reference frame $\epsilon = 1$ means that escapers have on average the progenitors angular velocity, whereas $\epsilon = 0$ corresponds to escapers having on average the progenitors orbital velocity). The equations of motion for stars that have escaped the progenitor are derived in Appendix~\ref{ap:qsg derivation} and it is important to note that these equations ignore the progenitor's mass as they are intended to describe the stripped stars' motion when the cluster potential experienced by the escapers is negligible compared to the galactic potential. The time-dependent angle from the centre of the potential of the ejected particle (relative to the progenitor) is given by

\begin{equation}\label{eq:QSG dtheta}
    \begin{array}{ll}
       \phi_1 (t) =  & -\frac{4-\gamma^2}{\gamma^2}\left(\Delta v_y + (1+\epsilon)\frac{\fesc r_{\rm J}}{R} V_{\rm c} \right)\frac{t}{R} \\
        & - \frac{2}{\gamma^3}(\gamma^2-2-2\epsilon)\frac{\fesc r_{\rm J}}{R}\sin\left(\gamma \Omega t\right) \\
        & + \frac{1}{\gamma^3} \frac{4 \Delta v_y}{V_{\rm c}} \sin\left(\gamma \Omega t\right) \\
        &- \frac{1}{\gamma^2} \frac{2\Delta v_x}{V_{\rm c}} \left(1-\cos\left(\gamma \Omega t\right)\right)  ,
    \end{array}
\end{equation}
where 
\begin{equation}
    \gamma^2 = 3+\frac{R^2}{V_{\rm c}^2}\partial_R^2 \Phi(R),
\end{equation}
is the ratio of epicyclic frequency to the angular frequency, $\Phi(R)$ is the spherical galactic potential, $t$ is the time since escape, and the negative sign in front of the equation means that stars ejected from the outer (inner) Lagrange point fall behind (move ahead) of the progenitor, as expected. 

The radial offset (that is, the displacement from the progenitor's orbital track in the direction of the galactic anti-centre) as a function of time is given by
\begin{equation}
    \begin{array}{ll}\label{eq:QSG dr}
        \displaystyle  \Delta r(t) = & \fesc r_{\rm J}\cos(\gamma \Omega t) + \frac{2R}{\gamma^2}
        \left(\frac{\Delta v_y}{V_{\rm c}} + (1+\epsilon)\frac{\fesc r_{\rm J}}{R}\right)\times \\
         & \left(1-\cos(\gamma \Omega t)\right) 
         +\frac{R \Delta v_x}{V_{\rm c}}\frac{\sin(\gamma \Omega t)}{\gamma}  ,
    \end{array}
\end{equation}
{where we note that $\Delta r(t=0)=\Delta x(t=0)$}.
The velocity in the $z$ direction simply tilts the orbital plane of the escaping star \citep*{Erkal_2016}
and the resulting motion perpendicular to the progenitor's orbital plane is given by
\begin{equation}\label{eq:QSG dz}
    \Delta z(t) = \Delta v_z \frac{R}{V_{\rm c}} \sin{\left(\frac{V_{\rm c}}{R} t\right)} .
\end{equation}

The density at a point along the stream is the product of the mass-loss rate and the $\phi_1$ distribution of that mass integrated from the start of stripping up until the observation time. This can be expressed in terms of the velocity distribution and used to map the density along the stream, $\rho(\phi_1, t)$, at all times 
\begin{equation}\label{eq:QSG rho(t)}
    \rho(\phi_1, t) =\frac{\gamma^2}{4-\gamma^2} \int_{0}^{t} \!  \frac{R}{t-t'} \, |\dot{M}(t')|\,P(\Delta v_y(\phi_1,t')) \, {\rm d}t' ,
\end{equation}
where $P(\Delta v_y(\phi_1,t))$ is the probability of the offset velocity required for a star to be at position $\phi_1$ at a given time. This equation sums up all possible stripping events up to the observation time and scales their contribution by the velocity distribution. By assuming a Gaussian distribution of offset velocities, the probability of a particular value is 
\begin{align}\label{eq:Pdvy}
\displaystyle    P(\Delta v_y) = \frac{1}{\sqrt{2\pi\sigma^2}} \, e^{-\frac{1}{2}\left(\frac{\Delta v_y}{\sigma}\right)^2} ,
\end{align}
where $\sigma$ is the velocity dispersion of a Plummer model \citep{Plummer_1911} at the escape radius,

\begin{equation}
    \sigma = \sqrt{\frac{GM}{6a}} \left[1+\left(\frac{\fesc r_{\rm J}}{a}\right)^2\right]^{-1/4}.
    \label{eq:sigmaesc}
\end{equation}
To get the ratio $\fesc r_{\rm J}/a$ we assume that the cluster fills the Roche radius, that is,  $r_{\rm h}/r_{\rm J}\simeq 0.15$ \citep{1961AnAp...24..369H}  and we note that $r_{\rm h} \simeq 1.305 \, a$ for Plummer's model. As demonstrated by equation~(\ref{eq:QSG dtheta}), there is a deterministic relation between $\phi_1$, $t$, and $\Delta v_y$, from which a time-dependent expression for $\Delta v_y$ is found by ignoring the oscillatory terms in equation~(\ref{eq:QSG dtheta}) to obtain the average motion of a star 
\begin{equation}\label{eq:QSG dtheta late times}
    \phi_1 (t) = - \frac{4-\gamma^2}{\gamma^2}\left(\Delta v_y + (1+\epsilon)\frac{\fesc r_{\rm J}}{R}V_{\rm c}\right)\frac{t}{R} ,
\end{equation}
and rearranging for $\Delta v_y$
\begin{equation}\label{eq:dvy}
    \Delta v_y(\phi_1,t) = -\frac{\gamma^2}{4-\gamma^2}\frac{\phi_1 R}{t} + (1+\epsilon)\frac{\fesc r_{\rm J}}{R}V_{\rm c} .
\end{equation}
By switching variables to $\tau = (t-t')^{-1}$ an expression that can be easily numerically evaluated is obtained,

\begin{equation}\label{eq:QSG rho(tau)}
    \rho(\phi_1, t) = \frac{\gamma^2}{4-\gamma^2}\int_{\tau_i}^{\tau_f}  \frac{R}{\tau} \,|\dot{M}(\tau)| \, P(\Delta v_y(\phi_1,\tau)) \, {\rm d}\tau .
\end{equation}
Equation~(\ref{eq:QSG rho(tau)}) can then be evaluated at regular intervals of $\phi_1$ at any point in time, before or after the progenitor has evaporated, to gain the 1-D density profile of the resulting stream. Hereafter, this (semi-)analytic model to quantify the stream growth is referred to as the QSG-AN (Quantifying Stream Growth - ANalytic) model. The model has two dimensionless parameters, $\epsilon$ and $\fesc$, that we will determine through a comparison to the $N$-body models in Section~\ref{ssec:comp}. $\epsilon$ and $\fesc$ relate the mean velocity of escapers to the escape radius and set the escape radius respectively. From this, and equation (\ref{eq:QSG dtheta late times}), it is clear that there is some degeneracy between them as both affect the mean drift velocity of escapers (that is,~the location of the centre of the $\Delta v_y$ distribution). However, $\fesc$ is the sole free parameter dictating the velocity dispersion in addition to the cluster mass. Throughout this work we use the analytic mass-loss rate from Section~\ref{ssec:mdot} as this gives us the smoothed case for a cluster with the given initial conditions removing the stochastic noise of dynamical ejections. However, one could also use the mass loss rate from numerical $N$-body simulations so long as it is corrected for the mass-loss due to stellar evolution.

\begin{figure*}
    \centering
    \includegraphics[width=.9\textwidth]{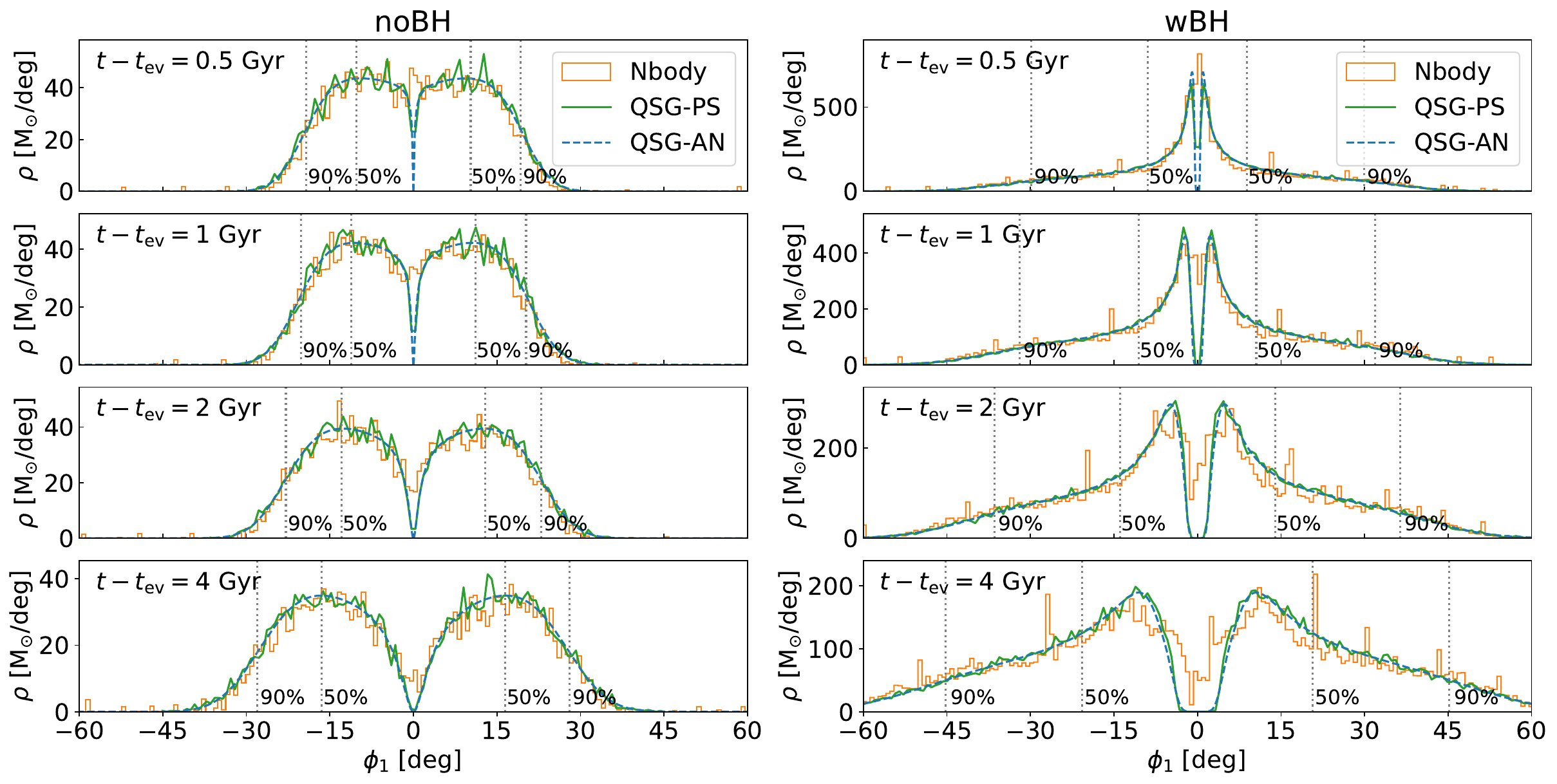}
    \caption{A comparison of the QSG-PS (green solid) and QSG-AN (blue dashed) models with the $N$-body simulations (orange) for the noBH model (\emph{left}) and wBH model (\emph{right}) at four times. From top to bottom these times are $0.5$, $1$, $2$, and $4~{\rm Gyr}$ after the progenitor has evaporated. The vertical dashed lines denote the $50^{\rm th}$ and $90^{\rm th}$ percentile by mass. As the QSG models predict the total mass distribution of a stream, the density profile of the $N$-body streams includes the contribution of white dwarfs and low-mass stars to provide a fair comparison. Due to the differing $M_{\rm i}$ and $\eta$, $0.5~{\rm Gyr}$ post-evaporation (top row) the peak linear density is over an order of magnitude greater in the wBH model than in the noBH model and remains a factor of $\sim5$ greater $4~{\rm Gyr}$ post-evaporation (bottom row). It is also seen that the $90^{\rm th}$ percentile is over twice the $50^{\rm th}$ percentile for the wBH model, whereas it is less than twice the $50^{\rm th}$ percentile in the noBH model.}
    \label{fig:QSG-AN comp}
\end{figure*}

\subsection{Particle spray method}
\label{ssec:pps}
QSG-AN does not capture all elements of a stream's structure, not only because it does not include the epicyclic over-densities due to the use of equation (\ref{eq:QSG dtheta late times}) which ignores the oscillatory terms, but because it only describes the one-dimensional density profile and therefore offers no insight in the stream offset from the orbit, nor the width. To describe these additional features of a stream we employ a Monte Carlo model using equations (\ref{eq:QSG dtheta}), (\ref{eq:QSG dr}), and (\ref{eq:QSG dz}) for motion along the stream, radially to the stream and perpendicular to the progenitor's orbital plane respectively, we refer to the resulting particle spray model as QSG-PS (Quantifying Stream Growth - Particle Spray). In QSG-PS a population of stars is generated, their escape times are calculated from equation~(\ref{eq:Mdot}) and their $\fesc r_{\rm J}$ are calculated  at their escape times using the mass of the cluster (equation~\ref{eq:M(t)}). Their velocity offset in each direction ($\Delta v_x$, $\Delta v_y$, $\Delta v_z$) is sampled from a Gaussian distribution centered on zero with a width equal to the velocity dispersion given by equation~(\ref{eq:sigmaesc}). Then, at any point in time prior- or post-evaporation those stars that have escaped the cluster can be selected and their positions calculated from equations~(\ref{eq:QSG dtheta}), (\ref{eq:QSG dr}), and (\ref{eq:QSG dz}). 

\begin{figure*}
    \centering
    \includegraphics[width=.9\textwidth]{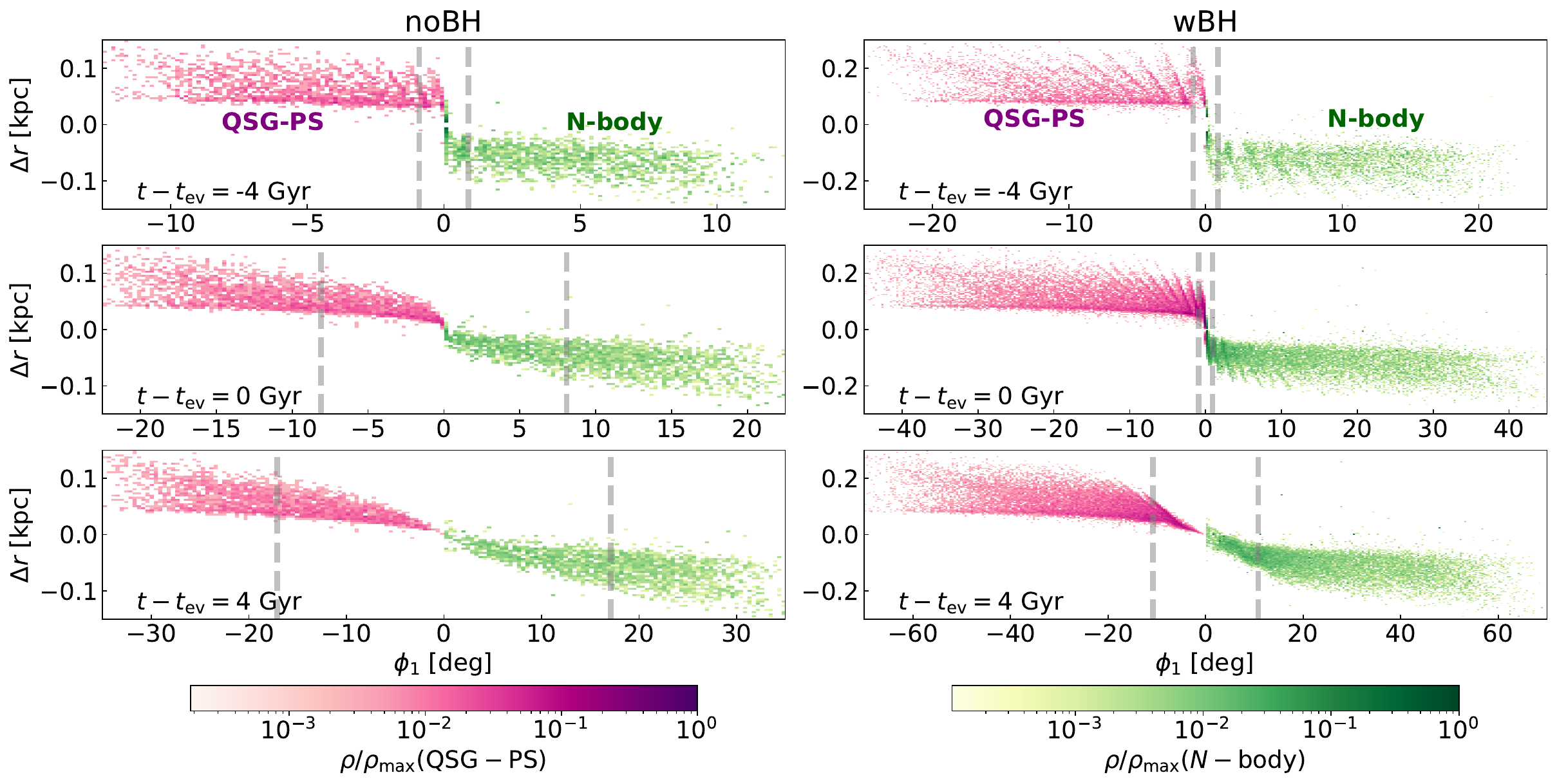}
    \caption{A comparison of QSG-PS (Red-Purple) with the $N$-body simulations (Yellow-Green) for the noBH models (left) and wBH models (right) for stream projections on the orbital plane. The density normalization, $\rho_{\rm max}$, is the maximum density across all the snapshots for both wBH and noBH models, hence why the noBH models never reach the maximum value. The leading and trailing tails are symmetric, such that no information is lost by showing one tail of each model, which we do here for ease of comparison. The peak of the linear density profile is denoted by the grey dashed line. 
    }
    \label{fig:QSG-PS comp}
\end{figure*}

\begin{figure*}
    \centering
    \includegraphics[width=.9\textwidth]{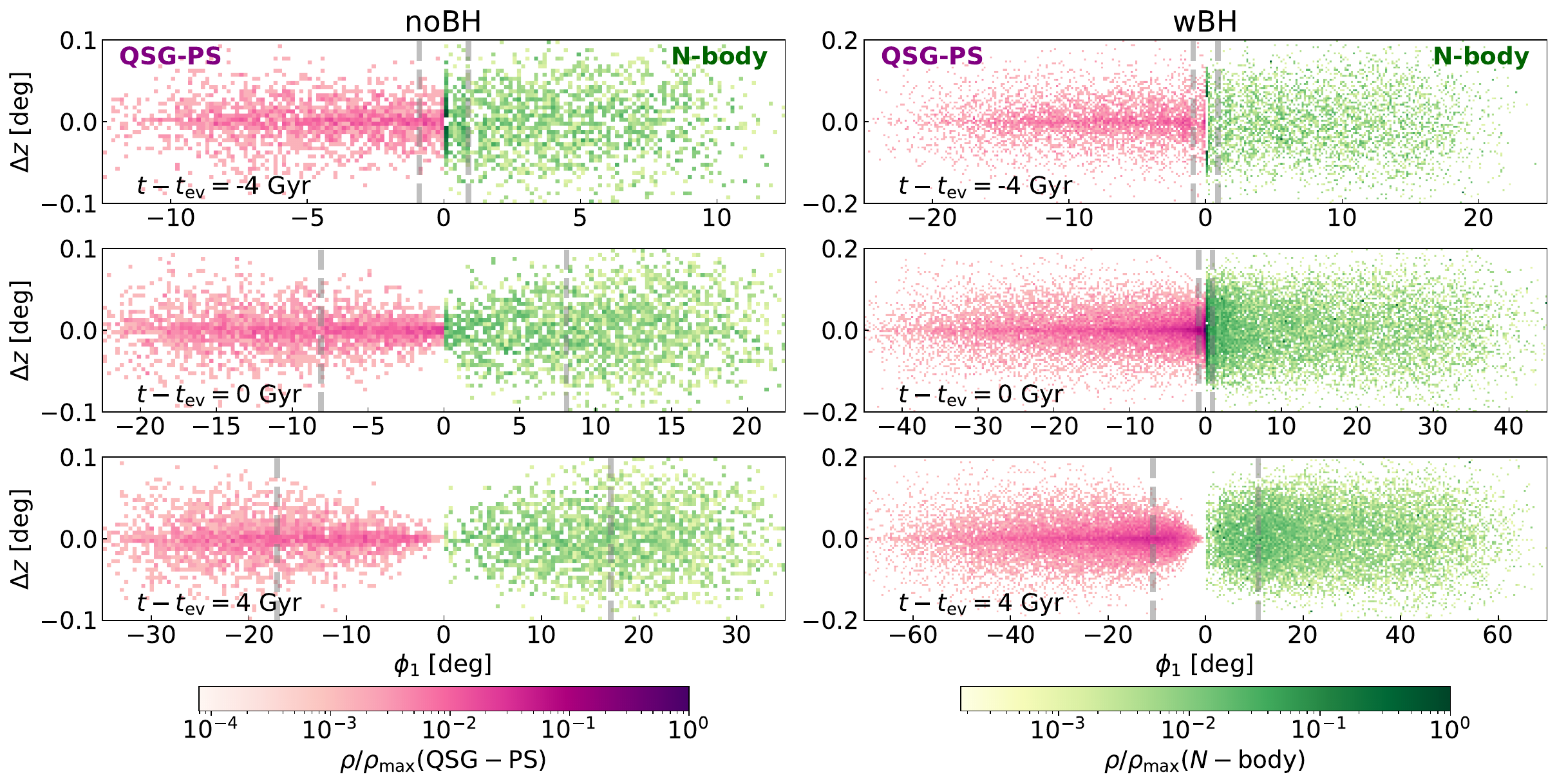}
    \caption{A comparison of QSG-PS (Red-Purple) with the $N$-body simulations (Yellow-Green) for the noBH models (left) and wBH models (right) for stream projections on the $\phi_1-z$ plane (that is as seen from the centre of the galaxy). Note that $\Delta z$ is equivalent to the commonly used $\Delta \phi_2$ coordinate for an observer at the Galactic centre. As in Fig. \ref{fig:QSG-PS comp}, the density normalization, $\rho_{\rm max}$, is the maximum density across all the snapshots for both wBH and noBH models and the leading and trailing tails are symmetric, such that no information is lost by showing one tail of each model. The peak of the linear density profile is denoted by the grey dashed line. 
    }
    \label{fig:QSG-PS comp yz}
\end{figure*}

\subsection{Comparison to $N$-body }
\label{ssec:comp}

To compare to the $N$-body models of Section~\ref{ssec:nbody}, we adopt an SIS potential for the galaxy with $\Vc = 220~{\rm km/s}$ as in the $N$-body models. We create stream models without BHs and with BHs, using both QSG-AN and QSG-PS, which we refer to as noBH-QSG and wBH-QSG, respectively. The models parameters are listed in Table \ref{tab:QSG ICs}. For all models we adopt circular orbits at $R=20~{\rm kpc}$ about the SIS potential. The analytic $M(t)$ and $\dot{M}(t)$ for these choices of parameters  are shown in Fig. \ref{fig:Mdot Mt} by the black dashed lines. In the QSG-PS model, $m_{*} = 0.36~\msun$ stars\footnote{Approximately the mean stellar mass for an old stellar population with a Kroupa IMF.}
are used such that the noBH-QSG has $N = 4622$ and wBH-QSG-PS model has $N=28638$, comparable to the $N$-body simulations. We approximate the mass in the BH population as a constant, $\Mbh = 150~\msun$, which is a reasonable approximation for $\Mbh(t)$ in the $N$-body simulation during the final $1~{\rm Gyr}$, in which the BHs have the strongest influence on the escape conditions of the stars (an increased $M_{\rm BH}$ would lead to greater differences between the wBH and noBH cases).

\begin{table}
    \centering
    \begin{tabular}{||c|c|c||}
        \hline
         & noBH-QSG & wBH-QSG \\
        \hline\hline
        Potential & SIS & SIS \\
        $R$ [kpc] & 20 & 20 \\
        \hline
        $V_{\rm c}$ [km/s] & 220 & 220 \\
        $M_{\rm i}$ [${\rm M}_{\odot}$] & 1664 & 10310 \\
        $\eta$ & 0.36 & -0.94 \\
        $\tev$ [Gyr] & 8.2 & 8.2\\
        $M_{\rm BH}$ [${\rm M}_{\odot}$] & 0 & 150 \\
        $N$ & 4622 & 28638 \\
        \hline
    \end{tabular}
    \caption{The parameters used in the noBH-QSG and wBH-QSG models.}
    \label{tab:QSG ICs}
\end{table}

We find the values for the two model parameters $\epsilon$ and $\fesc$ by comparing the distribution of stars in the $\phi_1$ and $\Delta r$ directions in the QSG-PS models and the density profile of the QSG-AN models to those of the $N$-body models. The resulting QSG-AN density profiles and QSG-PS streams are in very good agreement with the $N$-body simulations for $\epsilon\simeq 0.57$ and $\fesc \simeq 1.5$, as can be seen in Fig.~\ref{fig:QSG-AN comp} (QSG-AN) and Figs.~\ref{fig:QSG-PS comp} \&~\ref{fig:QSG-PS comp yz} (QSG-PS). It is these values that are used throughout the rest of this work in both QSG-AN and QSG-PS models and in both wBH and noBH cases. We stress that this is an effective model and as such one should not read too much into the meaning of $\epsilon$. Instead $\epsilon$ should be regarded as a parameter that can be obtained from a comparison to $N$-body models. However, for completeness, we note that $\epsilon < 1$ does indicate prograde in an inertial frame (retrograde in a co-rotating frame). 

As seen in Fig. \ref{fig:QSG-AN comp}, both QSG-AN and QSG-PS are both able to reproduce accurate linear density profiles for streams with and without BHs. However, after the progenitor has evaporated the density profile does deviate somewhat within the gap\footnote{We refer to the region between the two peaks centred at $\phi_1 = 0$ as `the gap', even though there are stars in this region.} at $\phi_1 = 0$. This is due to the idealised nature of our model assuming that even the final stars follow our prescription for the escape conditions. Despite this, they capture important features such as the peak of the density profile, size and shape of the wings, and the size of the gap remarkably well for such a simple model.

As is seen in Fig. \ref{fig:QSG-PS comp}, QSG-PS is able to reproduce the structure of the stream in the $\phi_1$-$\Delta r$ plane well in both wBH and noBH cases and both prior and post-evaporation. In particular, it captures well the length, width, offset from the progenitor's orbital track, and mass distribution along the stream. However, due to the simplifying assumptions and the idealised nature of the escape conditions there are some aspects where it diverges. In particular, because we assume a constant $M_{\rm BH}$ in the wBH-QSG-PS model there are fewer stars within the gap than in the wBH-Nbody model. In addition, during the final stages of evaporation we still assume that the stars follow our prescription for the escape conditions which may not be true and this manifests as the stream being very narrow near the progenitor's position post-evaporation, whereas the $N$-body models have a greater spread in $\Delta r$ near to the progenitor. This disparity is particularly noticeable in the wBH, 4\,Gyr (bottom, right) panel of Fig. \ref{fig:QSG-PS comp}. We see that this disparity between QSG-PS and $N$-body models is greatly reduced in the noBH models (see the bottom, left panel of Fig. \ref{fig:QSG-PS comp}), implying the need for a study of the escape conditions of stars in wBH GCs during the final stages of evaporation.

As seen in Fig. \ref{fig:QSG-PS comp yz}, QSG-PS does not fare as well in the $\phi_1$-$\Delta z$ plane. The simplistic nature of the QSG-PS model does not reproduce the diffuse nature of the stream in the $\phi_1\,-\,\Delta z$ plane. Instead, QSG-PS produces a much sharper over-density at the orbital plane, $\Delta z = 0$, which resembles a combination of a narrow and broad Gaussian. This is seen in other particle spray codes as well \citep[for example,][]{Gibbons+2014}. 

Throughout this work we compare the QSG models to the total stream in the $N$-body simulations (inclusive of white dwarfs and low-mass stars, which are not visible in observations) to facilitate a fair comparison. However, if one were to omit these objects to obtain the visible stream, then we note that the shape of the linear density profile is approximately unchanged and is just offset in normalisation. This is because low-mass stars are preferentially lost as the GC evolves towards energy equipartition and the white dwarfs are mainly lost at late times, with the two processes approximately balancing along the length of the stream.

The simple, fast, flexible nature of the QSG models makes them invaluable tools to explore the impact of the progenitor's properties on the structure of the resulting stream. With further refinement of the prescription of escape conditions to produce even more realistic streams, the QSG models have a wide range of application including constraining the possible parameter space of stream progenitors. In the next section we discuss the properties of the wBH and noBH streams in detail.

\section{The Impact of a Retained Black Hole Population}
\label{sec:impact}
From the models discussed in the previous section, we find that there are four main aspects of the stream's structure that differ due to the retained BH population: (1) The mass in the stream / inferred mass-loss rate; (2) the growth rate / stream length; (3) the shape of the central gap after the progenitor has evaporated and (4) the width and offset of the stream near the progenitor in the radial direction from the progenitor's orbital track soon after the progenitor has evaporated. The first three are because of differences in the  mass-loss rate and the fourth property is sensitive to the retained mass of the BH population. Below we discuss all four properties guided by the results from the models from the previous section. 

In this work, we consider only noBH and wBH streams of equal evaporation time, and as a result, they have differing initial masses. We posit that this is the most pertinent case to discuss because most observed GC streams have no progenitor ($\tev \lesssim 10~{\rm Gyr}$) and have not yet fully-phase mixed into the halo ($t-\tev \lesssim ~{\rm few}~{\rm Gyr}$), suggesting that there is a modest range of evaporation times, that is of order a few ${\rm Gyr}$. The alternate case of streams with the same initial mass, which is discussed in Appendix \ref{ap:same mi}, displays significant differences in stream length and density due to the differing evaporation times. Furthermore, in Appendix \ref{ap:same mi td} we briefly discuss the unphysical case of streams with the same initial mass and evaporation time, for completeness.

\subsection{Mass in the stream}
\label{ssec:massinstream}
\begin{figure}
    \centering
    \includegraphics[width=.45\textwidth]{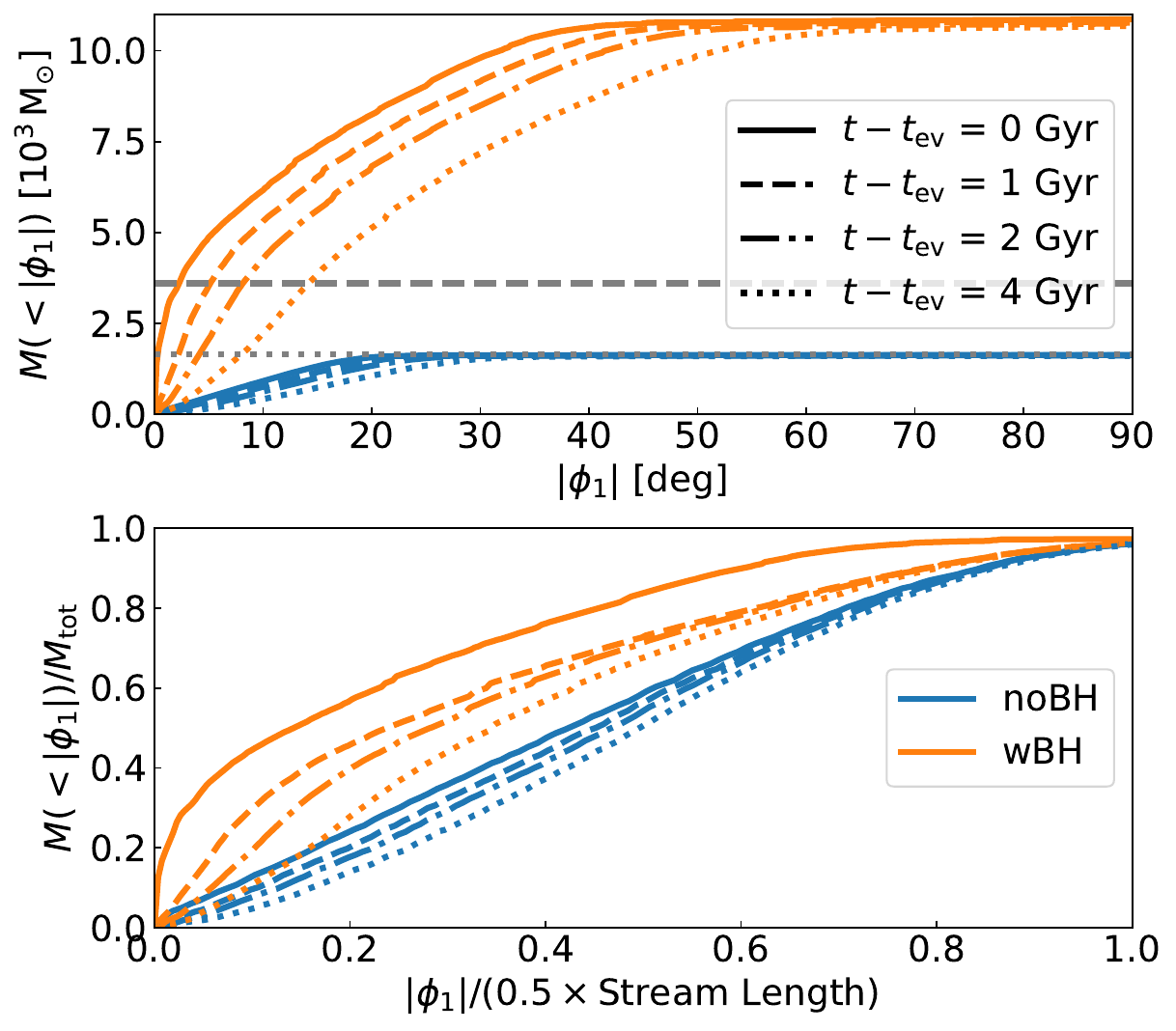}
    \caption{\emph{Top:} The cumulative mass as a function of the angular displacement from the progenitor along the stream for the noBH-Nbody (blue) and wBH-Nbody (orange) models at four times since evaporation (shown by the different line styles). The over-plotted grey dotted and dashed lines denote the $\Mi$ of the noBH-QSG stream and the maximum $\Mi$ of a noBH GC that can evaporate within a Hubble time, respectively. \emph{Bottom:} The cumulative mass fraction as a function of the fraction of the half stream length, where we define the stream length as twice the mass weighted $98^{\rm th}$ percentile of the stellar $|\phi_1|$ coordinates.}
    \label{fig:MassInStream}
\end{figure}

The top panel of Fig.~\ref{fig:MassInStream} displays the cumulative mass in the stream as a function of angular displacement from the progenitor's position for the noBH-Nbody and wBH-Nbody models at four times since evaporation ($t-\tev = 0,1,2,4~{\rm Gyr}$). These are compared to the initial mass of the noBH-Nbody model after stellar evolution ($M_{\rm i} = 1.66\times10^3~\msun$). It is observed that the wBH-Nbody model significantly exceeds the total mass of the noBH-Nbody model by $|\phi_1|\sim1\degree$ at $t-\tev=0~{\rm Gyr}$ and by $|\phi_1|\sim10\degree$ at $t-\tev=4~{\rm Gyr}$. In addition, Fig. \ref{fig:MassInStream} compares the cumulative mass in the stream to the mass, after stellar evolution, of a noBH GC that can evaporate on the same orbit within a Hubble time ($M_{\rm i} \sim 3.6\times10^3~\msun$) which is denoted by the dashed grey line. At $t-\tev=0~{\rm Gyr}$ the wBH stream exceeds this upper limit within $\sim 2.5\degree$ of the centre of the stream, and within $|\phi_1|\sim 15 \degree$ at $t - \tev = 4~{\rm Gyr}$. This demonstrates that we only need to observe a modest portion of a stream, if the orbit is well constrained, to deduce whether the density profile is consistent with the noBH case.

It is not just the total mass in the stream that differs but also the distribution of this mass (Figs. \ref{fig:QSG-AN comp} and \ref{fig:MassInStream}). As one would expect from the mass dependencies of the mass-loss rates of the wBH and noBH cases, the same fraction of the stream's mass is concentrated within a smaller fraction of the stream length in the wBH case than the noBH case. In the noBH-Nbody model the $\phi_1$ coordinate of the $90^{\rm th}$ percentile by mass is less than twice that of the $50^{\rm th}$ percentile, whereas in the wBH-Nbody model the $\phi_1$ coordinate of the $90^{\rm th}$ percentile is over twice that of the $50^{\rm th}$ percentile.

Since the gap grows after evaporation in both models, the enclosed mass at some distance of the wBH-Nbody model will eventually drop below $\Mi$ of the noBH-Nbody model. However, even at $t-\tev = 4~{\rm Gyr}$ the mass within $10\degree$ of the progenitor of the wBH-Nbody model exceeds the theoretical maximum for a BH-free GC on this orbit; this is owed not only to the greater initial mass of the wBH-Nbody model, but to the accelerating mass-loss rate resulting in the same fraction of the mass being concentrated closer to the progenitor's position as seen in Fig. \ref{fig:QSG-AN comp}. 

 In addition, the work of \citetalias{2023MNRAS.522.5340G} allows us to estimate the maximum expected density of a stream from a noBH progenitor on the same orbit with the same evaporation time ($\tev = 8.2~{\rm Gyr}$) from $\rho \propto \dot{M}/\overline{v}$. Taking the initial mass from section \ref{ssec:mdot} ($M_{\rm i} = 1.66\times10^3~\msun$) and the maximum mass-loss rate of this cluster on this orbit ($\dot{M}\simeq -0.3~\msun/{\rm Myr}$, from equation 1 of \citetalias{2023MNRAS.522.5340G}), we then need the mean drift velocity of a typical star, $\overline{v}$. This is found by differentiating equation (\ref{eq:QSG dtheta late times}) with respect to time 
\begin{equation}\label{eq:drift velocity}
    \dot{\phi}_{1}R = -\frac{4-\gamma^2}{\gamma^2}\left(\Delta v_y + (1+\epsilon)\frac{\fesc r_{\rm J}}{R}V_{\rm c}\right),
\end{equation}
obtaining\footnote{Note that the same expression can be obtained by differentiating and time averaging equation (\ref{eq:QSG dtheta}) over an integer number of epicycles and for a typical star ($\Delta v_y = 0$) this is the same expression as was derived in \citet{K_pper_2010} if $\fesc = \epsilon = 1$.} a mean drift velocity of $\overline{v}\simeq 0.8~{\rm pc/Myr}$. Dividing this mass-loss rate by the mean drift velocity and then by two, to account for mass escaping into the leading and trailing tail, gives an expected maximum density of $\rho \simeq 64~{{\rm M}_{\odot}/{\deg}}$ of a noBH stream on the same orbit. This density is $\sim 1.2$ times the maximum density in the top-left panel of Fig. \ref{fig:QSG-AN comp}, which is due to the fact that for the noBH model the maximum density in the tails occurs approximately when stripping begins. The wBH streams density exceeds this value out to $|\phi_1|\sim 30\degree$ for all panels of Fig. \ref{fig:QSG-AN comp}, except $t-\tev = 4~{\rm Gyr}$ where the central gap dips below it. This shows that the peak linear density is another useful metric in assessing the nature of the progenitor after it has evaporated and the centre of the stream is uncertain.

\subsection{Growth rate}
\label{ssec:growthrate}
The average growth rate of a stream can be approximated from equation (\ref{eq:drift velocity}) by considering a typical star that has the velocity offset equal to the mean of the Gaussian distribution upon escape ($\Delta v_x = \Delta v_y = 0$). This shows that the average speed at which a star moves along the stream is determined by $\rj$ at the time of release, with an adjustment to account for the offset velocity. This also shows that only the velocity in the $y$-direction leads to the bulk motion along the stream, whereas the velocity offsets in the $x$ and $z$ directions give rise to oscillations. For a circular orbit about a SIS potential this gives that a typical star travels along the stream with an average speed of $(1+\epsilon)\fesc r_{\rm J} \Omega$. In the simplest scenario, taking $\epsilon$ and $\fesc$ both to be unity, this agrees with \citet*{k2008}, showing that it is solely dependent on the mass of the progenitor, the progenitor's orbit, and the Galactic potential (through $\gamma$, as seen in equation \ref{eq:drift velocity}). Therefore, the wBH stream will grow approximately $1.8(M_{\rm i,wBH}/M_{\rm i,noBH})^{1/3}\simeq3.3$ times faster as the noBH stream, which can be clearly seen in Fig. \ref{fig:QSG-AN comp}.

Due to the low density at the extremes of the stellar streams,  uncertainties arise in formally defining the stream length, as well as resolving the ends of the streams from the background. Therefore, it is more pertinent to look at the peak of the linear density profile, which is the region that is most likely to be resolved in observations. The position of the peak of the linear density profile as a fraction of the stream length is determined by the mass-loss rate of the progenitor, while the $\phi_1$ coordinate is dictated by both the mass-loss rate and the mass of the progenitor cluster at the time that the stars that compose the peak escaped. 

When tidal stripping begins, the peak of the linear density profile is at the progenitor's location ($\phi_1 = 0$) for both wBH and noBH GCs. During the final stages of evaporation, the decelerating mass-loss rate of a noBH GC and differential streaming counterbalance one another to result in a linear density profile that is flat along approximately half the length of the tails, before declining to the end of the tails. However, for a wBH progenitor, the accelerating mass-loss rate and differential streaming compound one another to increase the magnitude of the density gradient along the stream, with the mass concentrated near the progenitor's position. Hence, in both noBH and wBH cases, the peak of the linear density profile is located at $\phi_1 = 0$ just before the progenitor has fully evaporated.

\begin{figure}
    \centering
    \includegraphics[width=.45\textwidth]{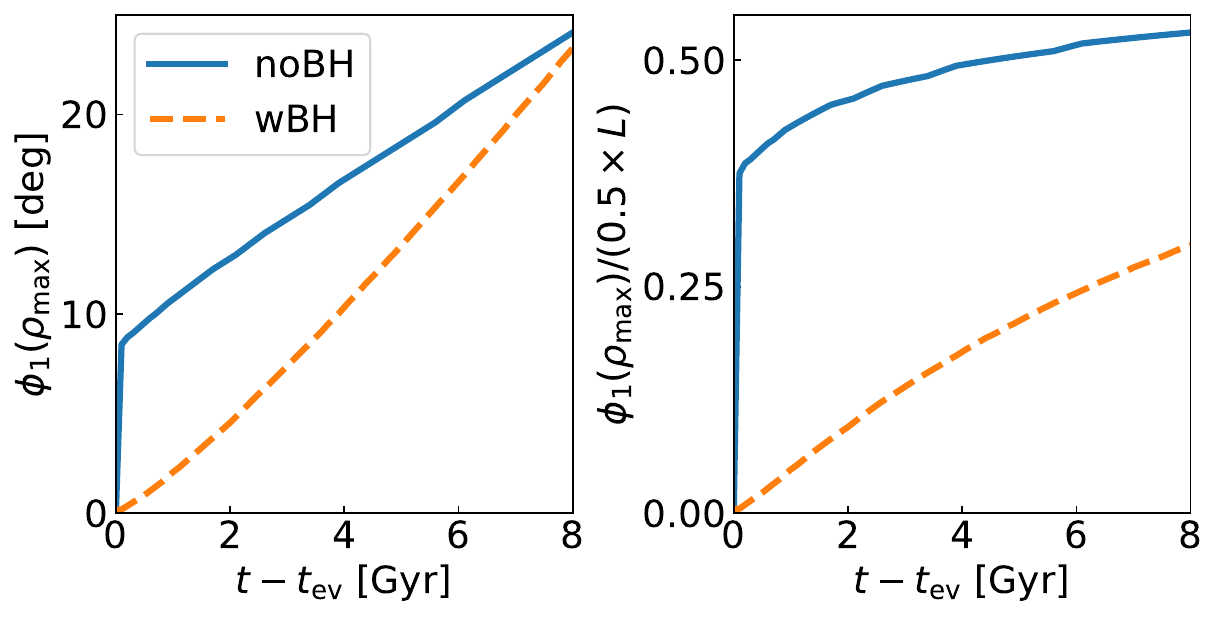}
    \caption{\emph{Left}: A comparison of   the position of the peak of the linear density profile as a function of time since evaporation. \emph{Right:} The position of the peak normalised by half the stream length (that is,~the distance from the centre of the stream, where $\phi_1 = 0$, to the $98^{\rm th}$ percentile by mass of $|\phi_1|$) as a function of time since evaporation.}
    \label{fig:PeakPos QSG-AN}
\end{figure}

Once the progenitor has evaporated, the propagation of stars away from the progenitor's position, and the velocity distribution of these stars leads to differential streaming occurring both at the extremes of the stream and at the centre of the stream ($\phi_1 = 0$). In the noBH case, this results in the peak of the linear density profile rapidly shifting along the tail until a quasi-equilibrium is reached approximately halfway from $\phi_1 = 0$ to the end of the stream, resulting in the formation of a gap. From this point forward, the peak remains approximately halfway from the centre of the stream to the end, asymptotically approaching $\phi_1(\rho_{\rm max})/(0.5L)\rightarrow0.6$ as $t-\tev\rightarrow\infty$, where $L$ is the stream length. Meanwhile, the stream is growing, resulting in the peak propagating at a near-constant rate, and differential streaming leads to the peak widening and flattening. This process can be seen in Fig. \ref{fig:PeakPos QSG-AN}, in which we see that at $t-\tev$ the peak is at $\phi_1 = 0\degree$, but at $t-\tev = 0.1~{\rm Gyr}$ it is located at $\phi_1 = 8.5\degree$ and, as seen in Fig. \ref{fig:QSG-AN comp}, at $t-\tev = 4~{\rm Gyr}$ the peak is located at $\phi_1 = 16.5\degree$ which is half the distance from the centre of the stream to the end.

In the wBH case, the propagation of stars away from the progenitor's position also results in the density peak, which was located at the progenitor's position at evaporation, propagating along the stream. Differential streaming not only causes the peak to widen and flatten, as in the noBH case, but for it to move along the length of the stream. As the peak moves to a greater fraction of the stream length, asymptotically approaching $\phi_1(\rho_{\rm max})/(0.5L)\rightarrow0.6$ as $t-\tev\rightarrow\infty$ (similar to what we found for the noBH case), the average velocity of the stars that compose the peak increases due to the greater cluster mass when they escaped, leading to the rate at which the peak propagates increasing. This process can be seen in Fig. \ref{fig:PeakPos QSG-AN}, in which we see that at $t-\tev = 0~{\rm Gyr}$ the peak is at $\phi_1 = 0\degree$,  at $t-\tev = 2~{\rm Gyr}$ it is located at $\phi_1 = 5\degree = 0.05L$, at $t-\tev = 4~{\rm Gyr}$ the peak is located at $\phi_1 = 10.4\degree = 0.08L$, and at $t-\tev = 8~{\rm Gyr}$ the peak is located at $\phi_1 = 23.3\degree = 0.15L$. 

As seen in the left-hand panel of Fig. \ref{fig:PeakPos QSG-AN}, for all noBH and wBH streams of equal evaporation time, there will exist a point at which the $\phi_1$ coordinate of the position of the peak for the noBH and wBH streams will coincide. The time since evaporation at which this intersection happens is dependent on the initial mass, orbit, the choice of free parameters ($\fesc$ and $\epsilon$), and $\eta$ (with a more positive $\eta$ resulting in a peak further along the stream and a more negative $\eta$ resulting in a peak closer to the progenitor's position post-evaporation). However, for the noBH-QSG and wBH-QSG models this occurs at $t-\tev\sim8.5~{\rm Gyr}$ (that is,~$\sim6.7~{\rm Gyr}$ in the future, if we assume the GC to were accreted $10~{\rm Gyr}$ ago). As seen in the right-hand panel of Fig. \ref{fig:PeakPos QSG-AN}, the position of the peak of the linear density profile as a fraction of the stream length is approximately four times greater in the noBH case than the wBH case, and remains approximately twice that of the wBH case at $t-\tev=8~{\rm Gyr}$.

\subsection{Central gap}
\label{ssec:centralgap}
\begin{figure}
    \centering
    \includegraphics[width=0.45\textwidth]{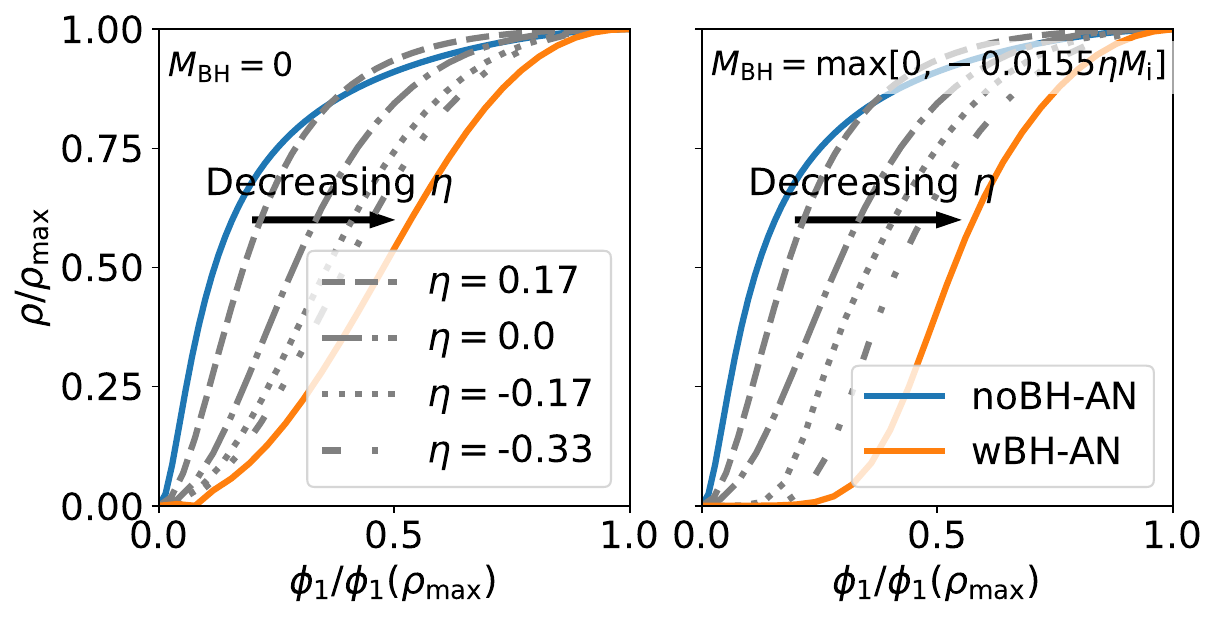}
    \caption{A plot of the impact of $\eta$ on the shape of the central gap at $\phi_1 = 0$ when $t-\tev = 2~{\rm Gyr}$ for GCs with $\tev = 8.2~{\rm Gyr}$. The limiting cases are noBH-QSG (blue) where $\eta=0.36$ and wBH-QSG (orange) where $\eta = -0.94$. \emph{Left:} $M_{\rm BH} = 0$ for all $\eta$. \emph{Right:} $M_{\rm BH} = 0$ for $\eta\geq0$ and $M_{\rm BH} = -0.0155\eta\Mi$ for $\eta<0$ such that $M_{\rm BH} \sim 150~\msun$ for wBH-QSG where $\eta = -0.94$ and $M_{\rm BH} = 32~\msun$ for the case of $\eta=-1/3$ and $\Mi = 6257~\msun$.}
    \label{fig:GapShape}
\end{figure}

The shape of the density profile of the gap from the progenitor's position at the centre to the peak of the linear density profile has an asymmetric ``S''-shape, as seen in Figs.~\ref{fig:QSG-AN comp} and \ref{fig:GapShape}. We define the shape of the linear density profile of the gap as the shape of the linear density profile from the centre to the peak of the linear density profile where $\phi_1$ and $\rho$ are normalised by the values of the peak (that is,~$\rho/\rho_{\rm max}$ and $\phi_1/\phi_1(\rho_{\rm max})$). The shape of this curve is primarily dictated by the mass-loss rate of the progenitor but it has a secondary dependence on the mass of the retained BH population. 

The left panel of Fig. \ref{fig:GapShape} displays the shape of the gap for streams of equal $\tev$ with differing values of $\eta$ and all have $M_{\rm BH} = 0$. The decelerating mass-loss rate of a noBH progenitor, as well as differential streaming, results in the density increasing from $\phi_1=0$ sharply and gently rounding off to the peak. The resulting shape is reminiscent of a function of the form $y=\sqrt[n]{x}$ or an incomplete beta function\footnote{Incomplete beta function of the form:\\ $I_x(a,b) = \Gamma(a+b)/[\Gamma(a)\Gamma(b)]\int_0^x\!t^{a-1}(1-t)^{b-1}\,{\rm d}t$} with $a\sim1,~b>1$. With increasing time since evaporation the widening and flattening of the peak leads to a gentler increase from $\phi_1 = 0$, replicated by an incomplete beta function where $a\sim1$ and $b$ is decreasing with time. Whereas, because the wBH progenitor's mass-loss rate was increasing as it evaporated, the shape of the linear density profile of the gap is dominated by the mass-loss just prior to evaporation, resulting in a more symmetrical shape that is reminiscent of the Gaussian profile one would expect from a single burst of mass-loss, or an incomplete beta function with $a\sim b > 1$. With time since evaporation the shape of the linear density profile of the gap of a wBH stream remains approximately constant.

The retained BH population has a secondary impact on the shape of the gap. As a noBH cluster approaches evaporation, the cluster mass tends to zero, leading to the Jacobi radius and, therefore, the mean velocity of the escaping stars to also tend to zero. This leads to mass being distributed throughout the gap and the linear density profile starting to increase from $\phi_1 = 0$. However, for a wBH progenitor approaching evaporation, the cluster mass tends to the mass of the retained BH population and, therefore, the average velocity of the final stars is given by equation \ref{eq:drift velocity} for $r_{\rm J}(M=M_{\rm BH})$. In the wBH-QSG models  $\rj\simeq14~{\rm pc}$ at $t-\tev = 0~{\rm Gyr}$, such that the final star escapes at a radial distance of $21~{\rm pc}$ from the centre of the cluster and, according to equation (\ref{eq:drift velocity}), has an average speed along the stream of $\sim 0.4~{\rm km/s}$ with respect to the progenitor assuming it has a velocity offset equal to the mean of the velocity distribution. In the wBH-Nbody model at evaporation, the average radial distance of stars within $50~{\rm pc}$ of the progenitor's position is $\sim 15.7~{\rm pc}$, such that according to equation (\ref{eq:drift velocity}) (taking $\epsilon = \fesc = 1$ for simplicity) these stars will have an average velocity of $\sim 0.35~{\rm km/s}$. This results in a gap with a flat bottom that expands with time since evaporation, as the final stars are better able to keep pace with the peak, and a shape reminiscent of a translated sigmoid function or an incomplete beta function with $a>b>1$. With increasing $M_{\rm BH}$, the more extensive the flat section of the linear density profile as the final stars have, on average, a higher velocity. This is an artifact that cannot be reproduced by a noBH stream because it is directly caused by the mass of the retained BH population.

Within QSG, the rate at which the gap expands or, due to our definition of the gap, equivalently the rate at which the density peaks propagate, can be well approximated from equation~(\ref{eq:drift velocity}) evaluated for the total stellar mass and BH mass contained within the gap, $M_{\rm gap}$, which we have defined as the region $|\phi_1|\leq\phi_1(\rho_{\rm max})$. This assumes that the peak of the linear density profile / end of the gap propagates with the mean velocity of the stars that compose that section of the stream and so we observe the rate at which the gap expands, $v_{\rm gap}$, to have a dependence on the orbit and galactic potential, as well as the mass contained within the gap, of the form
\begin{equation}
    v_{\rm gap} \propto \Omega^{1/3} M_{\rm gap}^{1/3}.
\end{equation}
For a noBH stream, the gap expands at near constant $v_{\rm gap}$ (Fig.~\ref{fig:PeakPos QSG-AN}), because $M_{\rm gap}$ is near constant post-evaporation. However, for a wBH stream,  $M_{\rm gap}$ is time varying, leading to $v_{\rm gap}$ increasing with time since evaporation, because the density peak moves to a greater fraction of the stream length post-evaporation. This variation in $v_{\rm gap}$ is only significant briefly after evaporation due to the weak dependence on $M_{\rm gap}$, such that in wBH-QSG $v_{\rm gap}$ can be considered approximately constant for $t-\tev\gtrsim3~{\rm Gyr}$, as demonstrated in Fig.~\ref{fig:PeakPos QSG-AN}.

When comparing the linear density profile of the gap of our idealised model to the $N$-body models (as in Fig. \ref{fig:QSG-AN comp}) we observe that, while the noBH-Nbody and noBH-QSG models are in good agreement and have a similar amount of mass retained within the gap, the centre of the gap in the wBH case is not devoid of mass as the QSG model predicts. Instead, as seen particularly in the $t-\tev = 4~{\rm Gyr}$ (bottom-left) panel of Fig. \ref{fig:QSG-AN comp}, the density peaks are not as sharp in wBH-Nbody as predicted by wBH-QSG because more stars exist within the region which wBH-QSG predicts to have no stars. This suggests that the motion of the final stars to escape the progenitor GC are no longer well approximated by the QSG equations of motion and perhaps that the assumption of a constant $M_{\rm BH}$ may be insufficient. This can be understood from the fact that during the final stages of evaporation the low-mass cluster is dominated of stars with energies well above the critical energy for escape (so-called `potential escapers', \citealt{2000MNRAS.318..753F}), which have average distances from the cluster center of  $\sim0.5\rj$ \citep*{10.1093/mnras/stw3309} from which they are able to escape because of their high energy. This breaks the explicit assumption of the QSG model that all stars escape through the Lagrange points.

\subsection{Stream width and radial offset}
To gain a sense of how the width and radial offset of a stream depend on the progenitor's properties, these quantities can be estimated from $\Delta r$. As the typical star escapes with $\Delta v_x = \Delta v_y = 0$, equation~(\ref{eq:QSG dr}) reduces to 
\begin{equation}\label{eq:QSG dr dvy dvx zero}
    \Delta r = \fesc r_{\rm J} \left[\cos(\gamma\Omega t) + \frac{2}{\gamma^2} (1+\epsilon) \left(1-\cos(\gamma\Omega t)\right)\right].
\end{equation}
The radial offset of the stream from the progenitor's orbital track can be approximated as the average radial displacement of a typical star over time
\begin{equation}\label{eq:mean dr}
    \overline{\Delta r} = \frac{2}{\gamma^2} (1+\epsilon) \fesc r_{\rm J},
\end{equation}
 The width of the stream in the radial direction, $w_x$, can be approximated as twice a typical star's mean displacement from $\overline{\Delta r}$,
\begin{equation}\label{eq:width x}
    w_x = 2\langle|\Delta r - \overline{\Delta r}|\rangle = \frac{4}{\pi}\left[\frac{2}{\gamma^2} (1+\epsilon) - 1\right] \fesc r_{\rm J},
\end{equation}
For a progenitor on a circular orbit about a SIS potential these simply become $\overline{\Delta r} = (1+\epsilon)\fesc r_{\rm J}$, $w_x = {4}{\pi}^{-1} \epsilon \fesc r_{\rm J}$. This shows that the width and radial offset are dependent on the escape radius. Note that these expressions are the same as can be obtained from the equations of motion of \citet{k2008} if $\epsilon = \fesc = 1$.

For the assumptions we have made of a circular orbit about a spherical potential, \citet{Erkal_2016} demonstrate that the width perpendicular to the orbital plane of the progenitor is determined by the spread in the stars orbital planes
\begin{equation}\label{eq:width z}
    w_z = \frac{\sigma}{\sqrt{2}\Omega} ,
\end{equation}
where $\sigma$ is the velocity dispersion at the escape radius given by equation (\ref{eq:sigmaesc}).

From equation~(\ref{eq:mean dr}), it is clear that our model predicts that the final stars in the wBH case should escape from $\sim21~{\rm pc}$ due to the retained BH population, whereas in the noBH case, as the final stars escape $r_{\rm J}$ tends to zero. This separation of $42~{\rm pc}$ between the centres of the tails should be detectable in direct observations. This offset which is present only in the wBH case, not the noBH case, is a feature that is solely due to the mass of the retained BH population.

When comparing the wBH-QSG and wBH-Nbody models distributions in the $x-y$ plane (see Fig. \ref{fig:QSG-PS comp}), we observe that the wBH-Nbody does not display such a clearly defined offset between the streams. This can be understood due to potential escapers no longer necessarily escaping through the Lagrange points. In addition, \citet{10.1093/mnras/stw3309} demonstrated that the velocity dispersion at the Jacobi radius is better fit by a $\sigma\propto M^{5/24}$ relation than the $\sigma\propto M^{1/3}$ relation of a Plummer model that we assume here, leading to a factor of two increase in the velocity dispersion at $M\simeq10^3~\msun$. All these factors contribute to the diffuse nature of the central portion of the stream. Nevertheless, as seen in the $t-\tev=4~{\rm Gyr}$ (bottom) panels of Fig. \ref{fig:QSG-PS comp}, the width of the stream in the central few degrees is wider in the wBH-Nbody model than the noBH-Nbody model even accounting for the differing initial masses, and this is owed to the mass of the retained BH population. However, real stellar streams do not exist in isolation, as our $N$-body models do. Perturbations from dark-matter subhalo (DMSH) fly-bys, disc shocks, or encounters with other GCs and dwarf galaxies can cause variations in the structure of the stream. Under the right circumstances, these could potentially cause a similar effect increasing the width in the central region.

Comparing the distributions in the $y-z$ plane, for the noBH-Nbody model we observe that post-evaporation the width in the $z$-direction, $w_{z}$, decreases significantly (by a factor of two at $t-\tev = 4~{\rm Gyr}$) from the location of the peak of the linear density profile to the centre. As can be seen from equation (\ref{eq:width z}), this is due to the decreasing cluster mass leading to the reduction in the velocity dispersion and, therefore, as you move to the centre the stream there is a smaller distribution of orbital planes. However, in the wBH-Nbody model we see that the width in the $z$-direction is more uniform along the length of the stream and $w_{z}$ at the centre is $\sim 0.9$ times $w_{z}$ at the peak of the linear density profile in the $t-\tev=4~{\rm Gyr}$ snapshot. This effect is due to the differing mass-loss rates.

\section{Discussion}
\label{sec:discussion}
The first three stream properties discussed in Sections~\ref{ssec:massinstream}, \ref{ssec:growthrate} and \ref{ssec:centralgap} formally depend on the mass-loss rate and not directly on the BHs, meaning that alternative scenarios that lead to similar mass-loss histories (such as a low initial cluster density, \citetalias{2021NatAs...5..957G}) might result in similar stream properties. This means that finding such tail properties is a necessary, but not sufficient, condition for the presence of BHs. However, \citetalias{2021NatAs...5..957G} show that the region in parameter space (low initial density and high initial mass) of clusters without BHs that have similarly high mass-loss rate is extremely small and this `fine-tuning' problem  makes the BH hypothesis a more likely scenario. We therefore conclude that if the tails suggest that the progenitor had a high mass-loss rate, the most likely interpretation is that the progenitor was rich in BHs.

One of the advantages of the QSG models that we present here is the speed at which they can generate realistic streams / density profiles, taking less than $0.01~{\rm s}$ for as single snapshot of wBH-QSG-AN/PS (including generating the initial conditions for QSG-PS). The speed of these models allows for quick exploration of parameter space compared to $N$-body models. These models are not just limited to investigating the impact of mass-loss rate on stream structure. The QSG models could potentially be used in conjunction with the equations of motion from \citet{erkal2015threephase} for stars after a DMSH fly-by to quickly narrow down the parameter space of the DMSH from the properties of the gap left behind.

The QSG-PS and QSG-AN models are able to quickly produce realistic streams and density profiles prior- or post-evaporation for both the noBH and wBH cases, as seen in Figs. \ref{fig:QSG-AN comp}, \ref{fig:QSG-PS comp}, \& \ref{fig:QSG-PS comp yz}. There are, however, a few aspects of the streams structure that QSG-PS does not capture well. Namely, the distribution in the $\phi_1$-$\Delta z$ plane and the distribution of central stars near the progenitor cluster post-evaporation which both require the implementation of more realistic escape conditions. Further refinement with the implementation of a more realistic velocity distribution of escapers such as that of \citet{10.1093/mnras/stw3309}, a time varying $M_{\rm BH}$, and a refined prescription of escape conditions following a study of stars escaping wBH GCs during the final stages of evaporation (which may call for a time varying $\fesc$) would greatly reduce these discrepancies making the insights gleaned from these models more reliable and impactful.

Our model is only valid for streams on circular orbits, but those on low eccentricity orbits can be approximated using a circular orbit of equal period with radius $R_{\rm T}$ and minimising the effect of the periodic stretching and compression due to the eccentricity by multiplying $\phi_1$ by $R(t)^2/R_{\rm T}^2$. For a Kepler orbit $R_{\rm T}$ is the semi-major axis and this approximation is within $1$ per cent ($10$ per cent) of the azimuthal period for orbits with eccentricities below $\sim 0.25$ ($0.75$) in a SIS. The expected orbit-averaged mass-loss rate for eccentric orbits can be calculated using an equivalent circular orbit with the same mass-loss rate (see \citetalias{2003MNRAS.340..227B}), which does not include any enhancement/reduction in mass-loss at peri-/apo-galacticon. This is important as GCs typically exist on eccentric orbits \citep{1997NewA....2..477O,1999ApJ...515...50V} and through this methodology not only can we project streams onto circular orbits to compare to the QSG models, but we can project the QSG models onto eccentric orbits allowing us to escape the confines of the circular orbit case.

The easiest approach to extend the model to eccentric orbits is with the QSG-PS model. The model parameters $\epsilon$ and $\fesc r_{\rm J}$ may need to be redetermined and may depend on eccentricity and/or the Galactic potential. A model for eccentric orbit can be deployed to directly infer the model parameters $\eta$, $\Mi$ from stream density profiles such as the ones presented in \cite{2022MNRAS.514.1757P}. Other improvements include the preferential escape of low-mass stars and variations in the stellar mass function which when combined with deep photometry of streams can provide additional constraints on the initial mass function (IMF) of GCs in addition to IMF constraints from mass modelling of GCs \citep{2023MNRAS.521.3991B, 2023MNRAS.522.5320D,2024MNRAS.529..331D}.
 
\section{Conclusions}
\label{sec:conclusions}
In this work we present a semi-analytical model of stream formation from star clusters on circular orbits and with time-dependent mass-loss rates and demonstrate that it produces streams that are in good agreement with $N$-body simulations of streams. The model has three free parameters that we determine from the comparison with the $N$-body models: (i) the mass dependency of the mass-loss rate, $\eta$, (ii) the mean distance of escape, $\fesc\rj$, and (iii) the relation between the escape radius and the centre of the velocity distribution, $\epsilon$. The best fit values are found to be $\epsilon \simeq 0.57$ and $\fesc \simeq 1.5$ for both values of $\eta$, our choice of initial mass, and Galactic orbit. We then compare the structure of streams resulting from progenitors that retain a stellar-mass black hole population (wBH) and those that do not (noBH). Retention of a stellar-mass BH population leads to streams that are more massive, have a peak closer to the progenitor location, have a narrower peak, and are more radially offset from the orbit. This is because wBH streams have approximately five times the mass of the equivalent noBH stream as the result of their accelerating mass-loss rate. It is also found that the shape of the central gap in the linear density profile is dependent on the mass-loss rate, and thereby the retained BH population. 

We also show that the limit on the mass of a noBH GC that can evaporate in a given time on a given orbit (\citetalias{2003MNRAS.340..227B}; \citetalias{2023MNRAS.522.5340G}) can be used to show that five of the seven streams (that are believed to have originated from GCs) included in \citet{2022MNRAS.514.1757P} have a mass that exceeds the initial mass of a noBH GC on an equivalent circular orbit that can evaporate in $10~{\rm Gyr}$ (see Fig.~\ref{fig:Mass Patrick Comp}). Not only does the orbit place a limit upon the mass in the stream, but also on the linear density, such that you do not necessarily need to observe the whole stream to find the mass within to be inconsistent with a noBH progenitor. This opens a new avenue to use the plethora of stellar streams without progenitors to learn about the black hole content of their evaporated progenitors.


\section*{Acknowledgements}
The authors thank Long Wang for discussions and help with {\sc PeTar} and Vasily Belokurov for helpful discussions. DR acknowledges support from the University of Southampton via the Mayflower studentship, the Erasmus+ programme of the European Union, and  thanks the ICCUB, where most of the work was conducted, for their hospitality. MG acknowledges support from the Ministry of Science and Innovation (EUR2020-112157, PID2021-125485NB-C22, CEX2019-000918-M funded by MCIN/AEI/10.13039/501100011033) and from AGAUR (SGR-2021-01069). DE acknowledges funding through ARC DP210100855.

\section*{Data Availability}
The data underlying this article will be shared on reasonable request to the corresponding authors. 


\bibliographystyle{mnras}
\bibliography{refs} 

\begin{thebibliography}{}
\makeatletter
\relax
\def\mn@urlcharsother{\let\do\@makeother \do\$\do\&\do\#\do\^\do\_\do\%\do\~}
\def\mn@doi{\begingroup\mn@urlcharsother \@ifnextchar [ {\mn@doi@} {\mn@doi@[]}}
\def\mn@doi@[#1]#2{\def\@tempa{#1}\ifx\@tempa\@empty \href {http://dx.doi.org/#2} {doi:#2}\else \href {http://dx.doi.org/#2} {#1}\fi \endgroup}
\def\mn@eprint#1#2{\mn@eprint@#1:#2::\@nil}
\def\mn@eprint@arXiv#1{\href {http://arxiv.org/abs/#1} {{\tt arXiv:#1}}}
\def\mn@eprint@dblp#1{\href {http://dblp.uni-trier.de/rec/bibtex/#1.xml} {dblp:#1}}
\def\mn@eprint@#1:#2:#3:#4\@nil{\def\@tempa {#1}\def\@tempb {#2}\def\@tempc {#3}\ifx \@tempc \@empty \let \@tempc \@tempb \let \@tempb \@tempa \fi \ifx \@tempb \@empty \def\@tempb {arXiv}\fi \@ifundefined {mn@eprint@\@tempb}{\@tempb:\@tempc}{\expandafter \expandafter \csname mn@eprint@\@tempb\endcsname \expandafter{\@tempc}}}

\bibitem[\protect\citeauthoryear{{Balbinot}, {Cabrera-Ziri}  \& {Lardo}}{{Balbinot} et~al.}{2022}]{2022MNRAS.515.5802B}
{Balbinot} E.,  {Cabrera-Ziri} I.,   {Lardo} C.,  2022, \mn@doi [\mnras] {10.1093/mnras/stac1953}, \href {https://ui.adsabs.harvard.edu/abs/2022MNRAS.515.5802B} {515, 5802}

\bibitem[\protect\citeauthoryear{{Banerjee} \& {Kroupa}}{{Banerjee} \& {Kroupa}}{2011}]{2011ApJ...741L..12B}
{Banerjee} S.,  {Kroupa} P.,  2011, \mn@doi [\apjl] {10.1088/2041-8205/741/1/L12}, \href {https://ui.adsabs.harvard.edu/abs/2011ApJ...741L..12B} {741, L12}

\bibitem[\protect\citeauthoryear{Banerjee, Belczynski, Fryer, Berczik, Hurley, Spurzem  \& Wang}{Banerjee et~al.}{2020}]{banerjee2020bse}
Banerjee S.,  Belczynski K.,  Fryer C.~L.,  Berczik P.,  Hurley J.~R.,  Spurzem R.,   Wang L.,  2020, Astronomy \& Astrophysics, 639, A41

\bibitem[\protect\citeauthoryear{{Baumgardt}}{{Baumgardt}}{2001}]{2001MNRAS.325.1323B}
{Baumgardt} H.,  2001, \mn@doi [\mnras] {10.1046/j.1365-8711.2001.04272.x}, \href {https://ui.adsabs.harvard.edu/abs/2001MNRAS.325.1323B} {325, 1323}

\bibitem[\protect\citeauthoryear{{Baumgardt} \& {Makino}}{{Baumgardt} \& {Makino}}{2003}]{2003MNRAS.340..227B}
{Baumgardt} H.,  {Makino} J.,  2003, \mn@doi [\mnras] {10.1046/j.1365-8711.2003.06286.x}, \href {http://adsabs.harvard.edu/cgi-bin/nph-bib_query?bibcode=2003MNRAS.340..227B&db_key=AST} {340, 227}

\bibitem[\protect\citeauthoryear{{Baumgardt}, {H{\'e}nault-Brunet}, {Dickson}  \& {Sollima}}{{Baumgardt} et~al.}{2023}]{2023MNRAS.521.3991B}
{Baumgardt} H.,  {H{\'e}nault-Brunet} V.,  {Dickson} N.,   {Sollima} A.,  2023, \mn@doi [\mnras] {10.1093/mnras/stad631}, \href {https://ui.adsabs.harvard.edu/abs/2023MNRAS.521.3991B} {521, 3991}

\bibitem[\protect\citeauthoryear{{Belokurov} et~al.,}{{Belokurov} et~al.}{2006}]{Belokurov+2006}
{Belokurov} V.,  et~al., 2006, \mn@doi [\apjl] {10.1086/504797}, \href {https://ui.adsabs.harvard.edu/abs/2006ApJ...642L.137B} {642, L137}

\bibitem[\protect\citeauthoryear{{Bernard} et~al.,}{{Bernard} et~al.}{2016}]{Bernard+2016}
{Bernard} E.~J.,  et~al., 2016, \mn@doi [\mnras] {10.1093/mnras/stw2134}, \href {https://ui.adsabs.harvard.edu/abs/2016MNRAS.463.1759B} {463, 1759}

\bibitem[\protect\citeauthoryear{{Bonaca} \& {Price-Whelan}}{{Bonaca} \& {Price-Whelan}}{2025}]{2025NewAR.10001713B}
{Bonaca} A.,  {Price-Whelan} A.~M.,  2025, \mn@doi [\nar] {10.1016/j.newar.2024.101713}, \href {https://ui.adsabs.harvard.edu/abs/2025NewAR.10001713B} {100, 101713}

\bibitem[\protect\citeauthoryear{Bonaca et~al.,}{Bonaca et~al.}{2020}]{Bonaca_2020}
Bonaca A.,  et~al., 2020, \mn@doi [The Astrophysical Journal] {10.3847/2041-8213/ab800c}, 892, L37

\bibitem[\protect\citeauthoryear{{Bonaca} et~al.,}{{Bonaca} et~al.}{2021}]{Bonaca+2021}
{Bonaca} A.,  et~al., 2021, \mn@doi [\apjl] {10.3847/2041-8213/abeaa9}, \href {https://ui.adsabs.harvard.edu/abs/2021ApJ...909L..26B} {909, L26}

\bibitem[\protect\citeauthoryear{{Bovy}}{{Bovy}}{2014}]{2014ApJ...795...95B}
{Bovy} J.,  2014, \mn@doi [\apj] {10.1088/0004-637X/795/1/95}, \href {https://ui.adsabs.harvard.edu/abs/2014ApJ...795...95B} {795, 95}

\bibitem[\protect\citeauthoryear{Bovy}{Bovy}{2015}]{bovy2015galpy}
Bovy J.,  2015, The Astrophysical Journal Supplement Series, 216, 29

\bibitem[\protect\citeauthoryear{{Bovy}, {Bahmanyar}, {Fritz}  \& {Kallivayalil}}{{Bovy} et~al.}{2016}]{Bovy+2016}
{Bovy} J.,  {Bahmanyar} A.,  {Fritz} T.~K.,   {Kallivayalil} N.,  2016, \mn@doi [\apj] {10.3847/1538-4357/833/1/31}, \href {https://ui.adsabs.harvard.edu/abs/2016ApJ...833...31B} {833, 31}

\bibitem[\protect\citeauthoryear{Breen \& Heggie}{Breen \& Heggie}{2013}]{10.1093/mnras/stt628}
Breen P.~G.,  Heggie D.~C.,  2013, \mn@doi [Monthly Notices of the Royal Astronomical Society] {10.1093/mnras/stt628}, 432, 2779

\bibitem[\protect\citeauthoryear{{Chen}, {Valluri}, {Gnedin}  \& {Ash}}{{Chen} et~al.}{2025}]{Chen+2024}
{Chen} Y.,  {Valluri} M.,  {Gnedin} O.~Y.,   {Ash} N.,  2025, \mn@doi [\apjs] {10.3847/1538-4365/ad9904}, \href {https://ui.adsabs.harvard.edu/abs/2025ApJS..276...32C} {276, 32}

\bibitem[\protect\citeauthoryear{Claydon, Gieles  \& Zocchi}{Claydon et~al.}{2017}]{10.1093/mnras/stw3309}
Claydon I.,  Gieles M.,   Zocchi A.,  2017, \mn@doi [Monthly Notices of the Royal Astronomical Society] {10.1093/mnras/stw3309}, 466, 3937

\bibitem[\protect\citeauthoryear{{Dickson}, {H{\'e}nault-Brunet}, {Baumgardt}, {Gieles}  \& {Smith}}{{Dickson} et~al.}{2023}]{2023MNRAS.522.5320D}
{Dickson} N.,  {H{\'e}nault-Brunet} V.,  {Baumgardt} H.,  {Gieles} M.,   {Smith} P.~J.,  2023, \mn@doi [\mnras] {10.1093/mnras/stad1254}, \href {https://ui.adsabs.harvard.edu/abs/2023MNRAS.522.5320D} {522, 5320}

\bibitem[\protect\citeauthoryear{{Dickson}, {Smith}, {H{\'e}nault-Brunet}, {Gieles}  \& {Baumgardt}}{{Dickson} et~al.}{2024}]{2024MNRAS.529..331D}
{Dickson} N.,  {Smith} P.~J.,  {H{\'e}nault-Brunet} V.,  {Gieles} M.,   {Baumgardt} H.,  2024, \mn@doi [\mnras] {10.1093/mnras/stae470}, \href {https://ui.adsabs.harvard.edu/abs/2024MNRAS.529..331D} {529, 331}

\bibitem[\protect\citeauthoryear{Erkal \& Belokurov}{Erkal \& Belokurov}{2015}]{erkal2015threephase}
Erkal D.,  Belokurov V.,  2015, Monthly Notices of the Royal Astronomical Society, 450, 1136

\bibitem[\protect\citeauthoryear{Erkal, Sanders  \& Belokurov}{Erkal et~al.}{2016}]{Erkal_2016}
Erkal D.,  Sanders J.~L.,   Belokurov V.,  2016, \mn@doi [Monthly Notices of the Royal Astronomical Society] {10.1093/mnras/stw1400}, 461, 1590

\bibitem[\protect\citeauthoryear{{Erkal} et~al.,}{{Erkal} et~al.}{2019}]{Erkal+2019}
{Erkal} D.,  et~al., 2019, \mn@doi [\mnras] {10.1093/mnras/stz1371}, \href {https://ui.adsabs.harvard.edu/abs/2019MNRAS.487.2685E} {487, 2685}

\bibitem[\protect\citeauthoryear{{Fardal}, {Huang}  \& {Weinberg}}{{Fardal} et~al.}{2015}]{Fardal+2015}
{Fardal} M.~A.,  {Huang} S.,   {Weinberg} M.~D.,  2015, \mn@doi [\mnras] {10.1093/mnras/stv1198}, \href {https://ui.adsabs.harvard.edu/abs/2015MNRAS.452..301F} {452, 301}

\bibitem[\protect\citeauthoryear{{Ferguson} et~al.,}{{Ferguson} et~al.}{2022}]{2022AJ....163...18F}
{Ferguson} P.~S.,  et~al., 2022, \mn@doi [\aj] {10.3847/1538-3881/ac3492}, \href {https://ui.adsabs.harvard.edu/abs/2022AJ....163...18F} {163, 18}

\bibitem[\protect\citeauthoryear{Fryer, Belczynski, Wiktorowicz, Dominik, Kalogera  \& Holz}{Fryer et~al.}{2012}]{Fryer_2012}
Fryer C.~L.,  Belczynski K.,  Wiktorowicz G.,  Dominik M.,  Kalogera V.,   Holz D.~E.,  2012, \mn@doi [The Astrophysical Journal] {10.1088/0004-637x/749/1/91}, 749, 91

\bibitem[\protect\citeauthoryear{{Fukushige} \& {Heggie}}{{Fukushige} \& {Heggie}}{2000}]{2000MNRAS.318..753F}
{Fukushige} T.,  {Heggie} D.~C.,  2000, \mn@doi [\mnras] {10.1046/j.1365-8711.2000.03811.x}, \href {https://ui.adsabs.harvard.edu/abs/2000MNRAS.318..753F} {318, 753}

\bibitem[\protect\citeauthoryear{{Gibbons}, {Belokurov}  \& {Evans}}{{Gibbons} et~al.}{2014}]{Gibbons+2014}
{Gibbons} S.~L.~J.,  {Belokurov} V.,   {Evans} N.~W.,  2014, \mn@doi [\mnras] {10.1093/mnras/stu1986}, \href {https://ui.adsabs.harvard.edu/abs/2014MNRAS.445.3788G} {445, 3788}

\bibitem[\protect\citeauthoryear{{Gieles} \& {Gnedin}}{{Gieles} \& {Gnedin}}{2023}]{2023MNRAS.522.5340G}
{Gieles} M.,  {Gnedin} O.~Y.,  2023, \mn@doi [\mnras] {10.1093/mnras/stad1287}, \href {https://ui.adsabs.harvard.edu/abs/2023MNRAS.522.5340G} {522, 5340 (GG23)}

\bibitem[\protect\citeauthoryear{{Gieles}, {Erkal}, {Antonini}, {Balbinot}  \& {Pe{\~n}arrubia}}{{Gieles} et~al.}{2021}]{2021NatAs...5..957G}
{Gieles} M.,  {Erkal} D.,  {Antonini} F.,  {Balbinot} E.,   {Pe{\~n}arrubia} J.,  2021, \mn@doi [Nature Astronomy] {10.1038/s41550-021-01392-2}, \href {https://ui.adsabs.harvard.edu/abs/2021NatAs...5..957G} {5, 957 (G21)}

\bibitem[\protect\citeauthoryear{{Giersz}, {Askar}, {Wang}, {Hypki}, {Leveque}  \& {Spurzem}}{{Giersz} et~al.}{2019}]{2019MNRAS.487.2412G}
{Giersz} M.,  {Askar} A.,  {Wang} L.,  {Hypki} A.,  {Leveque} A.,   {Spurzem} R.,  2019, \mn@doi [\mnras] {10.1093/mnras/stz1460}, \href {https://ui.adsabs.harvard.edu/abs/2019MNRAS.487.2412G} {487, 2412}

\bibitem[\protect\citeauthoryear{{Grillmair} \& {Dionatos}}{{Grillmair} \& {Dionatos}}{2006}]{GD1_disc}
{Grillmair} C.~J.,  {Dionatos} O.,  2006, \mn@doi [\apjl] {10.1086/505111}, \href {https://ui.adsabs.harvard.edu/abs/2006ApJ...643L..17G} {643, L17}

\bibitem[\protect\citeauthoryear{{H{\'e}non}}{{H{\'e}non}}{1961}]{1961AnAp...24..369H}
{H{\'e}non} M.,  1961, Ann. Astrophys., \href {http://adsabs.harvard.edu/abs/1961AnAp...24..369H} {24, 369}

\bibitem[\protect\citeauthoryear{Hurley, Pols  \& Tout}{Hurley et~al.}{2000}]{hurley2000SSE}
Hurley J.~R.,  Pols O.~R.,   Tout C.~A.,  2000, Monthly Notices of the Royal Astronomical Society, 315, 543

\bibitem[\protect\citeauthoryear{Hurley, Tout  \& Pols}{Hurley et~al.}{2002}]{hurley2002BSE}
Hurley J.~R.,  Tout C.~A.,   Pols O.~R.,  2002, Monthly Notices of the Royal Astronomical Society, 329, 897

\bibitem[\protect\citeauthoryear{{Ibata}, {Gilmore}  \& {Irwin}}{{Ibata} et~al.}{1994}]{1994Natur.370..194I}
{Ibata} R.~A.,  {Gilmore} G.,   {Irwin} M.~J.,  1994, \mn@doi [\nat] {10.1038/370194a0}, \href {https://ui.adsabs.harvard.edu/abs/1994Natur.370..194I} {370, 194}

\bibitem[\protect\citeauthoryear{{Ibata}, {Irwin}, {Lewis}, {Ferguson}  \& {Tanvir}}{{Ibata} et~al.}{2001}]{2001Natur.412...49I}
{Ibata} R.,  {Irwin} M.,  {Lewis} G.,  {Ferguson} A. M.~N.,   {Tanvir} N.,  2001, \mn@doi [\nat] {10.1038/35083506}, \href {https://ui.adsabs.harvard.edu/abs/2001Natur.412...49I} {412, 49}

\bibitem[\protect\citeauthoryear{Ibata, Malhan, Martin  \& Starkenburg}{Ibata et~al.}{2018}]{Ibata_2018}
Ibata R.~A.,  Malhan K.,  Martin N.~F.,   Starkenburg E.,  2018, \mn@doi [The Astrophysical Journal] {10.3847/1538-4357/aadba3}, 865, 85

\bibitem[\protect\citeauthoryear{{Ibata}, {Malhan}  \& {Martin}}{{Ibata} et~al.}{2019}]{2019ApJ...872..152I}
{Ibata} R.~A.,  {Malhan} K.,   {Martin} N.~F.,  2019, \mn@doi [\apj] {10.3847/1538-4357/ab0080}, \href {https://ui.adsabs.harvard.edu/abs/2019ApJ...872..152I} {872, 152}

\bibitem[\protect\citeauthoryear{{Koposov}, {Rix}  \& {Hogg}}{{Koposov} et~al.}{2010}]{2010ApJ...712..260K}
{Koposov} S.~E.,  {Rix} H.-W.,   {Hogg} D.~W.,  2010, \mn@doi [\apj] {10.1088/0004-637X/712/1/260}, \href {https://ui.adsabs.harvard.edu/abs/2010ApJ...712..260K} {712, 260}

\bibitem[\protect\citeauthoryear{{Koposov}, {Irwin}, {Belokurov}, {Gonzalez-Solares}, {Yoldas}, {Lewis}, {Metcalfe}  \& {Shanks}}{{Koposov} et~al.}{2014}]{Koposov+2014}
{Koposov} S.~E.,  {Irwin} M.,  {Belokurov} V.,  {Gonzalez-Solares} E.,  {Yoldas} A.~K.,  {Lewis} J.,  {Metcalfe} N.,   {Shanks} T.,  2014, \mn@doi [\mnras] {10.1093/mnrasl/slu060}, \href {https://ui.adsabs.harvard.edu/abs/2014MNRAS.442L..85K} {442, L85}

\bibitem[\protect\citeauthoryear{{Koposov} et~al.,}{{Koposov} et~al.}{2023}]{2023MNRAS.521.4936K}
{Koposov} S.~E.,  et~al., 2023, \mn@doi [\mnras] {10.1093/mnras/stad551}, \href {https://ui.adsabs.harvard.edu/abs/2023MNRAS.521.4936K} {521, 4936}

\bibitem[\protect\citeauthoryear{Kroupa}{Kroupa}{2001}]{kroupa2001IMF}
Kroupa P.,  2001, Monthly Notices of the Royal Astronomical Society, 322, 231

\bibitem[\protect\citeauthoryear{K{\"u}pper, Macleod  \& Heggie}{K{\"u}pper et~al.}{2008}]{k2008}
K{\"u}pper A.~H.,  Macleod A.,   Heggie D.~C.,  2008, Monthly Notices of the Royal Astronomical Society, 387, 1248

\bibitem[\protect\citeauthoryear{K{\"u}pper, Kroupa, Baumgardt  \& Heggie}{K{\"u}pper et~al.}{2010}]{K_pper_2010}
K{\"u}pper A. H.~W.,  Kroupa P.,  Baumgardt H.,   Heggie D.~C.,  2010, \mn@doi [Monthly Notices of the Royal Astronomical Society] {10.1111/j.1365-2966.2009.15690.x}, 401, 105–120

\bibitem[\protect\citeauthoryear{{K{\"u}pper}, {Balbinot}, {Bonaca}, {Johnston}, {Hogg}, {Kroupa}  \& {Santiago}}{{K{\"u}pper} et~al.}{2015}]{2015ApJ...803...80K}
{K{\"u}pper} A. H.~W.,  {Balbinot} E.,  {Bonaca} A.,  {Johnston} K.~V.,  {Hogg} D.~W.,  {Kroupa} P.,   {Santiago} B.~X.,  2015, \mn@doi [\apj] {10.1088/0004-637X/803/2/80}, \href {https://ui.adsabs.harvard.edu/abs/2015ApJ...803...80K} {803, 80}

\bibitem[\protect\citeauthoryear{{Kuzma}, {Da Costa}  \& {Mackey}}{{Kuzma} et~al.}{2018}]{2018MNRAS.473.2881K}
{Kuzma} P.~B.,  {Da Costa} G.~S.,   {Mackey} A.~D.,  2018, \mn@doi [\mnras] {10.1093/mnras/stx2353}, \href {http://adsabs.harvard.edu/abs/2018MNRAS.473.2881K} {473, 2881}

\bibitem[\protect\citeauthoryear{{Lamers}, {Baumgardt}  \& {Gieles}}{{Lamers} et~al.}{2010}]{2010MNRAS.409..305L}
{Lamers} H.~J.~G.~L.~M.,  {Baumgardt} H.,   {Gieles} M.,  2010, \mn@doi [\mnras] {10.1111/j.1365-2966.2010.17309.x}, \href {http://adsabs.harvard.edu/abs/2010MNRAS.409..305L} {409, 305}

\bibitem[\protect\citeauthoryear{{Li} et~al.,}{{Li} et~al.}{2022}]{Li+2022}
{Li} T.~S.,  et~al., 2022, \mn@doi [\apj] {10.3847/1538-4357/ac46d3}, \href {https://ui.adsabs.harvard.edu/abs/2022ApJ...928...30L} {928, 30}

\bibitem[\protect\citeauthoryear{{Lynden-Bell} \& {Lynden-Bell}}{{Lynden-Bell} \& {Lynden-Bell}}{1995}]{1995MNRAS.275..429L}
{Lynden-Bell} D.,  {Lynden-Bell} R.~M.,  1995, \mn@doi [\mnras] {10.1093/mnras/275.2.429}, \href {https://ui.adsabs.harvard.edu/abs/1995MNRAS.275..429L} {275, 429}

\bibitem[\protect\citeauthoryear{{Malhan}, {Ibata}  \& {Martin}}{{Malhan} et~al.}{2018}]{2018MNRAS.481.3442M}
{Malhan} K.,  {Ibata} R.~A.,   {Martin} N.~F.,  2018, \mn@doi [\mnras] {10.1093/mnras/sty2474}, \href {https://ui.adsabs.harvard.edu/abs/2018MNRAS.481.3442M} {481, 3442}

\bibitem[\protect\citeauthoryear{{Martin} et~al.,}{{Martin} et~al.}{2022}]{2022Natur.601...45M}
{Martin} N.~F.,  et~al., 2022, \mn@doi [\nat] {10.1038/s41586-021-04162-2}, \href {https://ui.adsabs.harvard.edu/abs/2022Natur.601...45M} {601, 45}

\bibitem[\protect\citeauthoryear{{Mart{\'\i}nez-Delgado} et~al.,}{{Mart{\'\i}nez-Delgado} et~al.}{2010}]{2010AJ....140..962M}
{Mart{\'\i}nez-Delgado} D.,  et~al., 2010, \mn@doi [\aj] {10.1088/0004-6256/140/4/962}, \href {https://ui.adsabs.harvard.edu/abs/2010AJ....140..962M} {140, 962}

\bibitem[\protect\citeauthoryear{{Mart{\'\i}nez-Delgado} et~al.,}{{Mart{\'\i}nez-Delgado} et~al.}{2023}]{2023A&A...671A.141M}
{Mart{\'\i}nez-Delgado} D.,  et~al., 2023, \mn@doi [\aap] {10.1051/0004-6361/202245011}, \href {https://ui.adsabs.harvard.edu/abs/2023A&A...671A.141M} {671, A141}

\bibitem[\protect\citeauthoryear{{Newberg}, {Willett}, {Yanny}  \& {Xu}}{{Newberg} et~al.}{2010}]{Newberg+2010}
{Newberg} H.~J.,  {Willett} B.~A.,  {Yanny} B.,   {Xu} Y.,  2010, \mn@doi [\apj] {10.1088/0004-637X/711/1/32}, \href {https://ui.adsabs.harvard.edu/abs/2010ApJ...711...32N} {711, 32}

\bibitem[\protect\citeauthoryear{{Odenkirchen}, {Brosche}, {Geffert}  \& {Tucholke}}{{Odenkirchen} et~al.}{1997}]{1997NewA....2..477O}
{Odenkirchen} M.,  {Brosche} P.,  {Geffert} M.,   {Tucholke} H.~J.,  1997, \mn@doi [\na] {10.1016/S1384-1076(97)00035-3}, \href {https://ui.adsabs.harvard.edu/abs/1997NewA....2..477O} {2, 477}

\bibitem[\protect\citeauthoryear{{Odenkirchen} et~al.,}{{Odenkirchen} et~al.}{2001}]{2001ApJ...548L.165O}
{Odenkirchen} M.,  et~al., 2001, \mn@doi [\apjl] {10.1086/319095}, \href {http://adsabs.harvard.edu/abs/2001ApJ...548L.165O} {548, L165}

\bibitem[\protect\citeauthoryear{{Patrick}, {Koposov}  \& {Walker}}{{Patrick} et~al.}{2022}]{2022MNRAS.514.1757P}
{Patrick} J.~M.,  {Koposov} S.~E.,   {Walker} M.~G.,  2022, \mn@doi [\mnras] {10.1093/mnras/stac1478}, \href {https://ui.adsabs.harvard.edu/abs/2022MNRAS.514.1757P} {514, 1757}

\bibitem[\protect\citeauthoryear{{Pavlík}, {Jeřábková}, {Kroupa}  \& {Baumgardt}}{{Pavlík} et~al.}{2018}]{Pavlik2018BHretention}
{Pavlík} V.,  {Jeřábková} T.,  {Kroupa} P.,   {Baumgardt} H.,  2018, \mn@doi [A&A] {10.1051/0004-6361/201832919}, 617, A69

\bibitem[\protect\citeauthoryear{Plummer}{Plummer}{1911}]{Plummer_1911}
Plummer H.~C.,  1911, \mn@doi [Monthly Notices of the Royal Astronomical Society] {10.1093/mnras/71.5.460}, 71, 460

\bibitem[\protect\citeauthoryear{{Sanders}}{{Sanders}}{2014}]{2014MNRAS.443..423S}
{Sanders} J.~L.,  2014, \mn@doi [\mnras] {10.1093/mnras/stu1159}, \href {https://ui.adsabs.harvard.edu/abs/2014MNRAS.443..423S} {443, 423}

\bibitem[\protect\citeauthoryear{{Shipp} et~al.,}{{Shipp} et~al.}{2018}]{2018ApJ...862..114S}
{Shipp} N.,  et~al., 2018, \mn@doi [\apj] {10.3847/1538-4357/aacdab}, \href {https://ui.adsabs.harvard.edu/abs/2018ApJ...862..114S} {862, 114}

\bibitem[\protect\citeauthoryear{{Wang}}{{Wang}}{2020}]{2020MNRAS.491.2413W}
{Wang} L.,  2020, \mn@doi [\mnras] {10.1093/mnras/stz3179}, \href {https://ui.adsabs.harvard.edu/abs/2020MNRAS.491.2413W} {491, 2413}

\bibitem[\protect\citeauthoryear{Wang, Iwasawa, Nitadori  \& Makino}{Wang et~al.}{2020}]{wang2020petar}
Wang L.,  Iwasawa M.,  Nitadori K.,   Makino J.,  2020, Monthly Notices of the Royal Astronomical Society, 497, 536

\bibitem[\protect\citeauthoryear{{Wang}, {Gieles}, {Baumgardt}, {Li}, {Pang}  \& {Tang}}{{Wang} et~al.}{2024}]{2024MNRAS.527.7495W}
{Wang} L.,  {Gieles} M.,  {Baumgardt} H.,  {Li} C.,  {Pang} X.,   {Tang} B.,  2024, \mn@doi [\mnras] {10.1093/mnras/stad3657}, \href {https://ui.adsabs.harvard.edu/abs/2024MNRAS.527.7495W} {527, 7495}

\bibitem[\protect\citeauthoryear{{de Boer}, {Erkal}  \& {Gieles}}{{de Boer} et~al.}{2020}]{2020MNRAS.494.5315D}
{de Boer} T.~J.~L.,  {Erkal} D.,   {Gieles} M.,  2020, \mn@doi [\mnras] {10.1093/mnras/staa917}, \href {https://ui.adsabs.harvard.edu/abs/2020MNRAS.494.5315D} {494, 5315}

\bibitem[\protect\citeauthoryear{{van den Bosch}, {Lewis}, {Lake}  \& {Stadel}}{{van den Bosch} et~al.}{1999}]{1999ApJ...515...50V}
{van den Bosch} F.~C.,  {Lewis} G.~F.,  {Lake} G.,   {Stadel} J.,  1999, \mn@doi [\apj] {10.1086/307023}, \href {https://ui.adsabs.harvard.edu/abs/1999ApJ...515...50V} {515, 50}

\makeatother
\end{thebibliography}

\appendix
\section{Cluster mass evolution with and without BHs}
\label{app:mt}
We use the expression for the evaporation time for clusters with different amounts of BHs of \citetalias{2023MNRAS.522.5340G} (their equation 6) 
\begin{equation}
    \tev\propto \frac{\Mi^x}{y \dot{M}_{\rm ref}\Omega_{\rm tid}}.
\end{equation}
This equation has four parameters: (i) $x$ sets the relation between the evaporation time and the initial mass (after stellar evolution), which is fixed to $x=2/3$; (ii) $y$ sets the evolution of the escape rate which depends on the BH content: constant for $y=1$, accelerating for $y>1$ (wBH) or decelerating for $y<1$ (noBH); (iii) $\dot{M}_{\rm ref}$ is the mass-loss rate at fixed reference mass ($2\times 10^5~\msun$), and (iv) $\Omega_{\rm tid}$ is a measure of the strength of the tidal field, which we take to be that of a singular isothermal sphere (SIS), $\Omega_{\rm tid}=\sqrt{2}V_{\rm c}/R_{\rm eff}$, where $V_{\rm c}$ is the circular velocity. For the three cases shown in Fig.~\ref{fig:Mass Patrick Comp} these are $(y,\mdotref)=(2/3, -30\,\msun/\myr)$ for `noBH' (approximately the \citetalias{2003MNRAS.340..227B} results and the parameters as used in \citetalias{2023MNRAS.522.5340G}'s model without BHs, see their equation 1), $(4/3, -45\,\msun/\myr)$ for `wBH median' (the parameters as used in \citetalias{2023MNRAS.522.5340G}'s model which accounts for the effect of BHs, see their equation 21) and $(2, -95\,\msun/\myr)$ for `wBH max' which corresponds to \citetalias{2023MNRAS.522.5340G}'s lowest density $N$-body models where the effect of BHs on the mass-loss rate is maximal. Throughout this paper we use noBH to refer to GCs that at some point early in their evolution reach $f_{\rm BH} < f_{\rm BH,crit}$, causing all  BHs to be dynamically ejected, leading to GCs evolving similarly to GCs that are initially BH free \citep{10.1093/mnras/stt628}. We use wBH to refer to GCs that have $f_{\rm BH} > f_{\rm BH,crit}$ and will eventually become BH dominated  \citep[$f_{\rm BH}\rightarrow1$,][]{2011ApJ...741L..12B}. \citet{10.1093/mnras/stt628} found $f_{\rm BH,crit} = 0.1$ for their two component models, whereas \citetalias{2023MNRAS.522.5340G} found a lower value of $f_{\rm BH,crit} \sim 0.025$ from their $N$-body models. This has implications for GCs, as star clusters of metallicity $Z \sim 0.0001 ~ - ~ 0.001$ (${\rm [Fe/H]} \sim -2.1 ~ $ to $ ~ -1.1$) are expected to have an initial BH fraction of $f_{\rm BH,0}\sim 0.05$ (see \citealt{banerjee2020bse} their figure 7, \citetalias{2023MNRAS.522.5340G} their figure 4), meaning that dense clusters  eject all their BHs early on, while lower density clusters can keep BHs until today.

\section{Deriving the Equation of Motion}\label{ap:qsg derivation}
Here a derivation of the equations of motion of escaped stars in the cluster centred frame from section \ref{ssec:stream growth} is presented. Using the common substitution $u = 1/R$ and expanding the equations of motion at leading order gives

\begin{equation}\label{QSG deriv: eq of motion}
    \frac{{\rm d}^2 \Delta u}{{\rm d}\theta^2} + \gamma^2\Delta u = -2 u_0 \left(\frac{\Delta v_y}{V_{\rm c}} + \frac{\fesc r_{\rm J}}{R}\right)
\end{equation}
where $\theta$ is the azimuthal angle about the Galactic centre and 
\begin{equation}
    \gamma^2 = 3+\frac{R^2}{V_{\rm c}^2}\partial_R^2 \Phi.
\end{equation}
The corresponding initial conditions of $u$ are given by 
$\Delta u(0) = -\fesc r_{\rm J}/R^2$
 and 
$\partial_{\theta}\Delta u(0) = -u \Delta v_x/V_{\rm c}$. 
Giving a solution to equation~(\ref{QSG deriv: eq of motion}) of
\begin{equation}
 \displaystyle    
    \begin{array}{ll}
        \Delta u = & -\frac{\fesc r_{\rm J}}{R^2} \cos(\gamma \theta) - \frac{2u_0}{\gamma^2} \left[ \frac{\Delta v_y}{V_{\rm c}} + \frac{\fesc r_{\rm J}}{R}\right]\left[1-\cos(\gamma \theta)\right] \\
        & - \frac{u_0 \Delta v_x}{V_{\rm c}} \frac{\sin(\gamma \theta)}{\gamma}.
    \end{array}
\end{equation}
Switching back from $u$ to $R$
\begin{equation}
    \begin{array}{ll}
        \Delta R = &  \fesc r_{\rm J} \cos(\gamma \theta) - \frac{2R}{\gamma^2} \left[ \frac{\Delta v_y}{V_{\rm c}} + \frac{\fesc r_{\rm J}}{R}\right]\left[1-\cos(\gamma \theta)\right]\\
        & - \frac{R \Delta v_x}{V_{\rm c}} \frac{\sin(\gamma \theta)}{\gamma}.
   \end{array}
\end{equation}
Using conservation of angular momentum
\begin{equation}\label{eq:conserv. ang. mom.}
    L_z = R^2 \dot{\theta} ,
\end{equation}
$\dot{\theta}$ can be expressed as
\begin{equation}
    \begin{array}{ll}
        \dot{\theta} = & \frac{V_{\rm c}}{R}\left[1-\left(\frac{\Delta v_y}{V_{\rm c}} + \frac{\fesc r_{\rm J}}{R}\right) \left(\frac{4-\gamma^2}{\gamma^2}\right) \right. \\
        &\left. -2 \left(\frac{\fesc r_{\rm J}}{R} \frac{\gamma^2 -2}{\gamma^2} - \frac{2}{\gamma^2} \frac{\Delta v_y}{V_{\rm c}}\right) \cos(\gamma \theta) -2 \frac{\Delta v_x}{V_{\rm c}} \frac{\sin(\gamma\theta)}{\gamma}\right].
   \end{array}
\end{equation}
Integrating over $t$, and approximating $\theta$ as $V_{\rm c}t/R$ at leading order, the angular displacement relative to the progenitor can be expressed as
\begin{equation}
    \begin{array}{ll}
        \phi_1(t,\Delta v_x,\Delta v_y) = & -\frac{4-\gamma^2}{\gamma^2} \left(\frac{\Delta v_y}{V_{\rm c}} + \frac{\fesc r_{\rm J}}{R}\right) \frac{V_{\rm c} t}{R} \\
        & - \frac{2}{\gamma^3}\left((\gamma^2 -2)\frac{\fesc r_{\rm J}}{R} - \frac{2\Delta v_y}{V_{\rm c}}\right) \sin\left(\frac{\gamma V_{\rm c} t}{R}\right) \\
        & - \frac{2\Delta v_x}{V_{\rm c}} \frac{1-\cos\left(\frac{\gamma V_{\rm c} t}{R}\right)}{\gamma^2}.
    \end{array}
\end{equation}
This expression gives the angular displacement of a particle from the progenitor as a function of time and the escape conditions. Since the Lagrange points move at the same angular rate as the progenitor (for a circular orbit), the mean velocity at the Lagrange point can be related to the radial offset through a free parameter $\epsilon$
\begin{equation}
    \frac{\Delta v_y}{V_{\rm c}} \rightarrow \epsilon \frac{\fesc r_{\rm J}}{R} + \frac{\Delta v_y}{V_{\rm c}},
\end{equation}
where $\Delta v_y$ is now the random component of the velocity with a mean of zero. $\epsilon = 1$ for a constant angular rate. Re-writing $\Delta r$ and $\phi_1$

\begin{equation}
    \begin{array}{ll}
        \phi_1 (t) =  & -\frac{4-\gamma^2}{\gamma^2}\left(\Delta v_y + (1+\epsilon)\frac{\fesc r_{\rm J}}{R} V_{\rm c} \right)\frac{t}{R} \\
        & - \frac{2}{\gamma^3}(\gamma^2-2-2\epsilon)\frac{\fesc r_{\rm J}}{R}\sin\left(\gamma \frac{V_{\rm c}}{R}t\right) \\
        & + \frac{1}{\gamma^3} \frac{4 \Delta v_y}{V_{\rm c}} \sin\left(\gamma \frac{V_{\rm c}}{R}t\right) \\
        & - \frac{1}{\gamma^2} \frac{2\Delta v_x}{V_{\rm c}} \left(1-\cos\left(\gamma \frac{V_{\rm c}}{R}t\right)\right)  ,
    \end{array}
\end{equation}

\begin{equation}
    \begin{array}{ll}
        \Delta r(t) = & \fesc r_{\rm J}\cos(\gamma\phi_1) + \frac{2R}{\gamma^2}
        \left(\frac{\Delta v_y}{V_{\rm c}} + (1+\epsilon)\frac{\fesc r_{\rm J}}{R}\right)\times \\
        & \left(1-\cos(\gamma\phi_1)\right) 
        +\frac{R \Delta v_x}{V_{\rm c}}\frac{\sin(\gamma\phi_1)}{\gamma}  ,
    \end{array}
\end{equation}

where we have changed $\Delta R$ to $\Delta r$ to signify that, in the context of this work, we are considering it as the displacement from the progenitor's orbital track in a cluster centered frame rather than the change in galactocentric radius in the galactocentric frame, of course the two are equivalent in this model restricted to circular orbits.

\section{Streams of Equal initial mass}\label{ap:same mi}
Here we present figures the same as Figs.~\ref{fig:QSG-AN comp} and \ref{fig:QSG-PS comp} but for wBH and noBH streams of equal initial mass. For these, we calculate the evaporation times from equation 6 of \citetalias{2023MNRAS.522.5340G} and we assume a moderate value for $\eta$ ($-0.33$) and $M_{\rm BH}$ ($50~\msun$) in the wBH case. Previously when comparing to the $N$-body model we used $m_{*} = 0.36~\msun$ to try and mimic the results, however, here we use $m_{*} = 0.01~\msun$ to be able to observe all the substructure that is encapsulated within the model, we stress that this does not alter the results of the model in any way, it simply allows for a higher resolution plot.

\begin{table}
    \centering
    \begin{tabular}{||c|c|c||}
        \hline
        Equal $M_{\rm i}$ & noBH-QSG-C & wBH-QSG-C \\
        \hline\hline
        Potential & SIS & SIS \\
        $R$ [kpc] & 20 & 20 \\
        $V_{\rm c}$ [km/s] & 220 & 220 \\
        $M_i$ [${\rm M}_{\odot}$] & 3000 & 3000 \\
        $\eta$ & 0.33 & -0.33 \\
        $\tev$ [Gyr] & 12.0 & 4.0\\
        $M_{\rm BH}$ [${\rm M}_{\odot}$] & 0 & 50 \\
        \hline
    \end{tabular}
    \caption{The parameters used in the noBH-QSG-C and wBH-QSG-C models. These models have equal initial mass and the evaporation time is calculated from eq. 6 of \citetalias{2023MNRAS.522.5340G}.}
    \label{tab:QSG SAME MI ICs}
\end{table}

\begin{figure*}
    \centering
    \includegraphics[width=.9\textwidth]{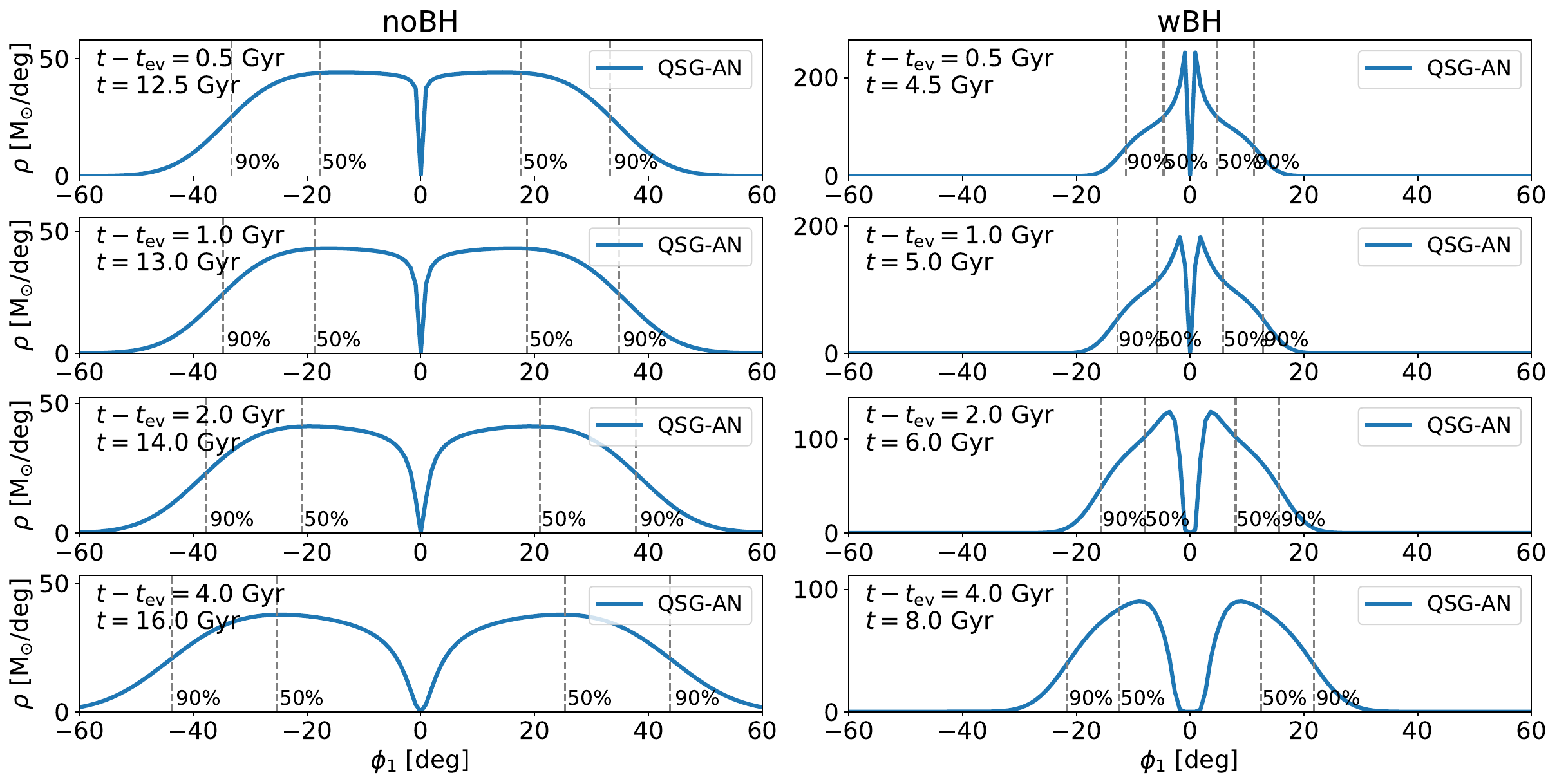}
    \caption{The linear density profiles for the noBH-QSG-C (\emph{left}) and wBH-QSG-C (\emph{right}) streams which have the same initial masses on the same orbits.}
    \label{fig:SAME MI density}
\end{figure*}

Fig.~\ref{fig:SAME MI density} displays the linear density profile of the noBH-QSG-C and wBH-QSG-C streams which have the same initial mass. We see that, despite the streams having the same initial mass, the resulting differing evaporation times lead to clear differences in the streams that make them easily differentiable. The wBH stream is approximately one third the length of the noBH stream and has a maximum density over four times the maximum density of the noBH stream $0.5~{\rm Gyr}$ after evaporation. Even at $4~{\rm Gyr}$ after evaporation, the wBH stream is approximately half the length and has a maximum density twice that of the noBH stream. These peak densities of wBH-QSG-C are above the theoretical maximum density of the noBH-QSG-C model of $\sim63~\msun/{\rm deg}$ and remains above this value up to $t-\tev\sim8~{\rm Gyr}$.

\begin{figure*}
    \centering
    \includegraphics[width=.9\textwidth]{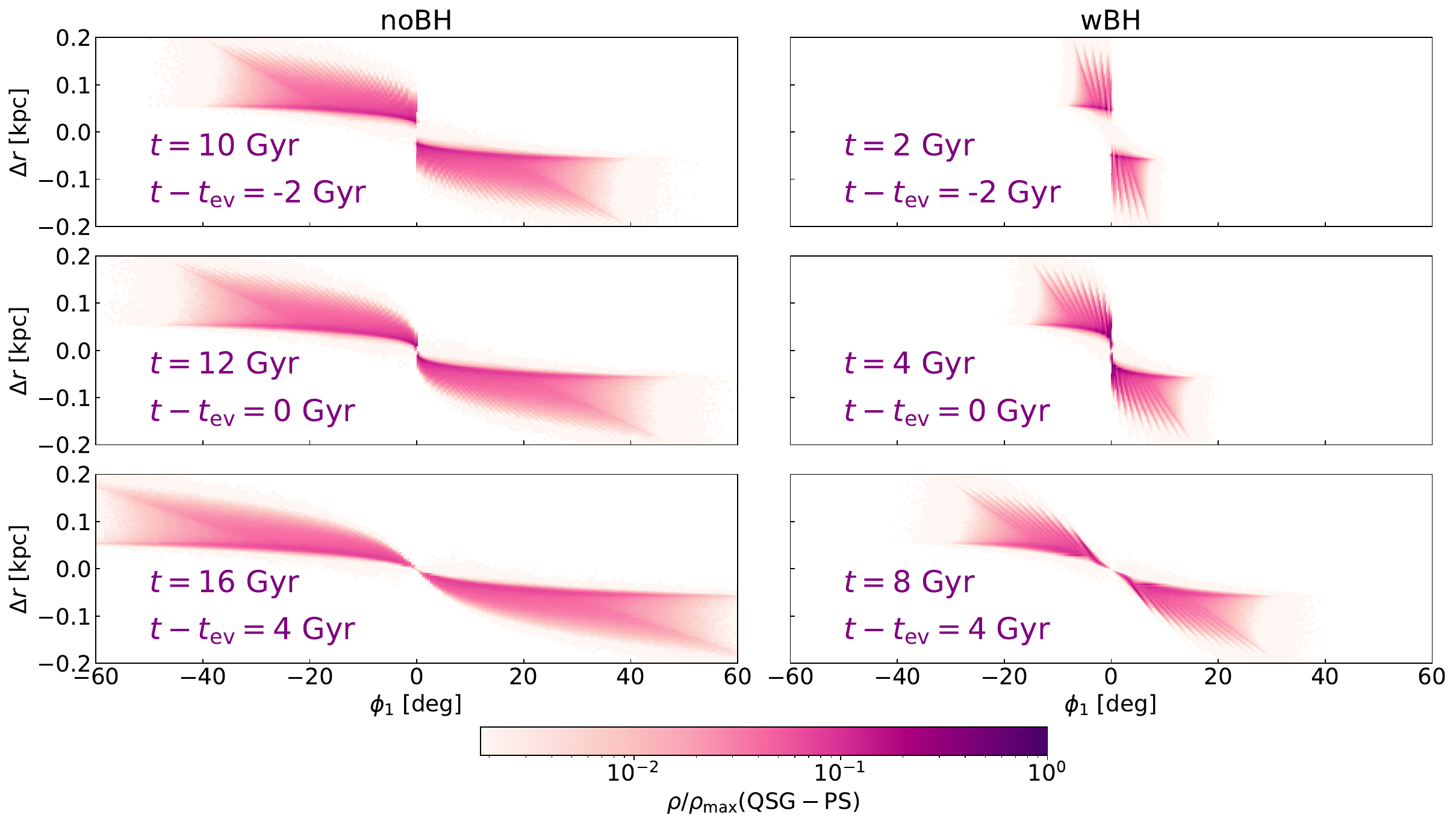}
    \caption{A plot of the density distribution of the noBH-QSG-C (\emph{left}) and wBH-QSG-C (\emph{right}) in the $x-y$ plane at $t-\tev = -2,0,4~{\rm Gyr}$ (from top to bottom). }
    \label{fig:SAME MI XY}
\end{figure*}

Fig.~\ref{fig:SAME MI XY} displays the mass distribution in the $x-y$ plane and as expected from equations~\ref{eq:mean dr} \& \ref{eq:width x} these streams have the same average width and radial offset due them having the same initial mass. However, due to the differing mass-dependencies of the mass loss rate, within the central portion of the stream, the wBH stream will be wider at the same fraction of the stream length. This is because the accelerating mass loss-rate ensures that the cluster mass will have been greater when these stars were released. However, this is a minor effect due to the weak mass dependency and, therefore, is unlikely to be useful in determining the nature of an observed stream.

In this scenario the key metrics to differentiate between a wBH and a noBH stream are the stream length and the location and magnitude of the peak of the linear density profile.

\section{Streams of Equal initial mass and evaporation time}\label{ap:same mi td}
Here, for completeness, we present unphysical models of noBH-QSG-D and wBH-QSG-D streams with the same $\Mi$ and $\tev$. We stress that it is unphysical to have a wBH stream and a noBH stream of equal mass and evaporation time, where the noBH GC has $\eta\sim1/3$.

\citetalias{2021NatAs...5..957G}, demonstrated that it is possible to have a noBH GC with a mass-loss rate akin to a wBH GC. However, the high-mass and low initial density required occupy a very small area of parameter space leading to the ``fine-tuning'' problem described in \citetalias{2021NatAs...5..957G}. In this case, the noBH progenitor would have an accelerating mass-loss rate ($\eta\lesssim -1/3$ rather than the canonical noBH value of $\eta\sim1/3$) and the only difference between the two streams would be the width of the stream at $\phi_1 \sim 0$ at $t\sim \tev$.

\begin{table}
    \centering
    \begin{tabular}{||c|c|c||}
        \hline
        Equal $\Mi$ and $\tev$ & noBH-QSG-D & wBH-QSG-D \\
        \hline\hline
        Potential & SIS & SIS \\
        $R$ [kpc] & 20 & 20 \\
        $V_{\rm c}$ [km/s] & 220 & 220 \\
        $M_i$ [${\rm M}_{\odot}$] & 3000 & 3000 \\
        $\eta$ & 0.33 & -0.33 \\
        $\tev$ [Gyr] & 4.0 & 4.0\\
        $M_{\rm BH}$ [${\rm M}_{\odot}$] & 0 & 50 \\
        \hline
    \end{tabular}
    \caption{The parameters used in the noBH-QSG-D and wBH-QSG-D models. These unphysical models have equal initial masses and evaporation times.}
    \label{tab:QSG SAME MI TD ICs}
\end{table}

\begin{figure*}
    \centering
    \includegraphics[width=.9\textwidth]{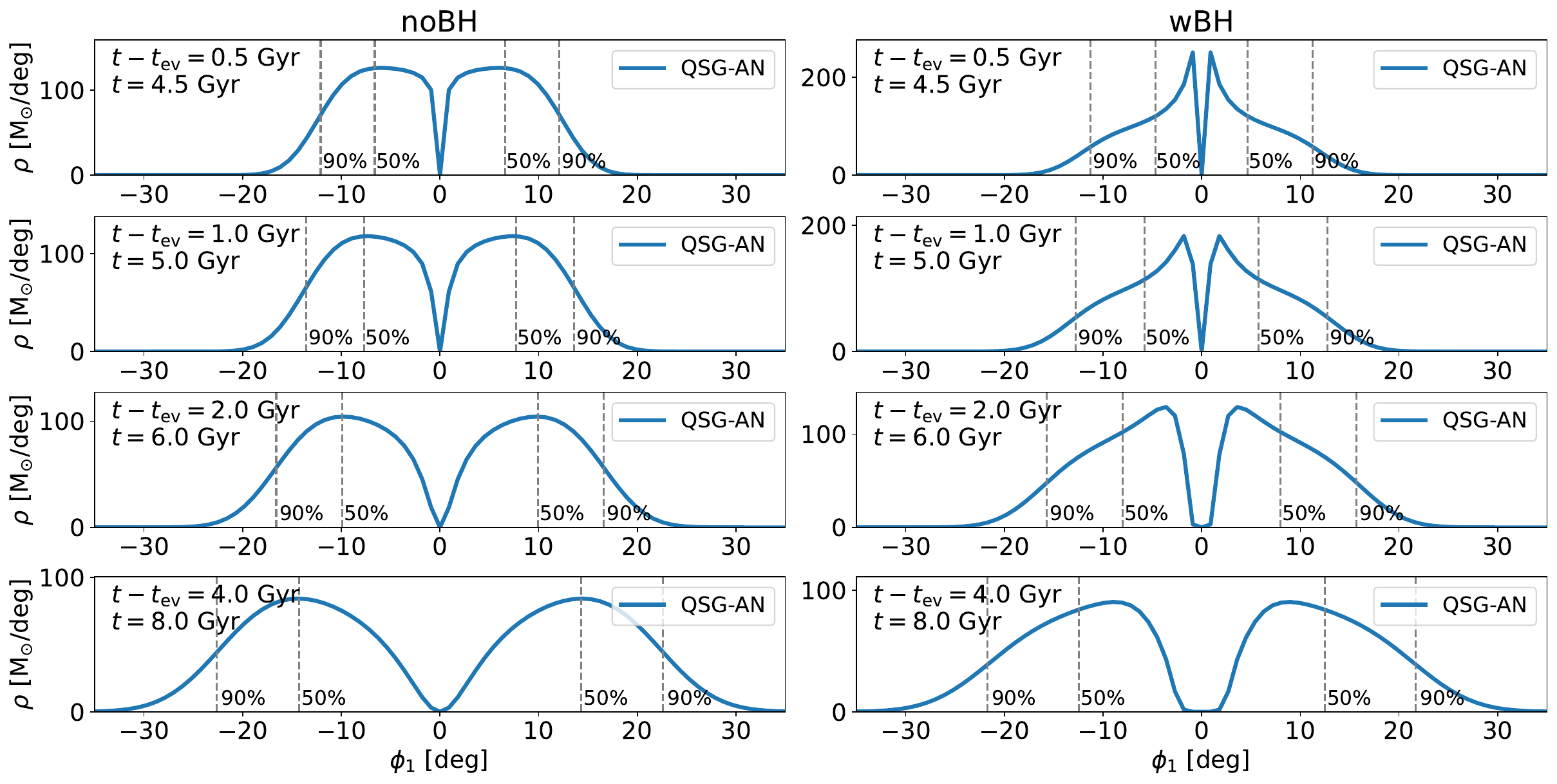}
    \caption{The linear density profiles for the noBH-QSG-D (\emph{left}) and wBH-QSG-D (\emph{right}) streams which have the same initial mass and evaporation time, as well as being on the same orbits.}
    \label{fig:SAME MI TD DENSITY}
\end{figure*}

Fig.~\ref{fig:SAME MI TD DENSITY} displays the linear density profile of the streams of equal initial mass and evaporation time. It is observed that, due to the mass dependency of the mass-loss rate, the maximum density of the wBH-QSG-D is over 1.5 times the density of the noBH-QSG-D stream at $t-\tev = 0.5~{\rm Gyr}$, and that the mass is concentrated closer to the progenitor. Fig.~\ref{fig:SAME MI TD DENSITY} demonstrates the differing gap morphologies that result from the differing mass-loss rates and the mass of the retained BH population. With time since evaporation the density profiles become more alike, such that at $4~{\rm Gyr}$ after evaporation they resemble one another, with a rounded shape and similar maximum densities. However, the position of the peak of the linear density profile, both the $\phi_1$ coordinate and as a fraction of the stream length, are different with the noBH peak's position being approximately twice that of the wBH stream.

\begin{figure*}
    \centering
    \includegraphics[width=.9\textwidth]{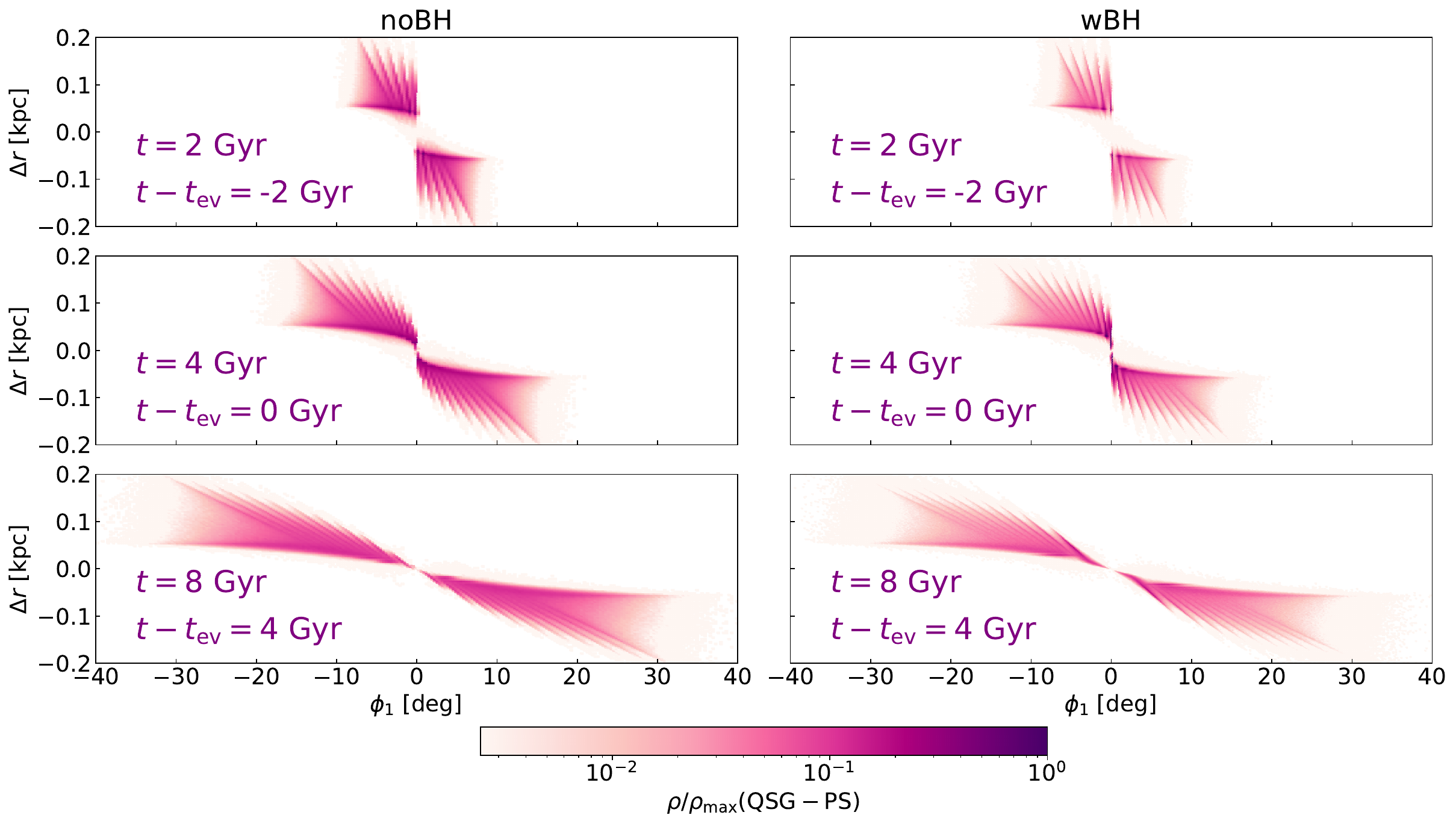}
    \caption{A plot of the density distribution of the noBH-QSG-D (\emph{left}) and wBH-QSG-D (\emph{right}) in the $x-y$ plane at $t-\tev = -2,0,4~{\rm Gyr}$ (from top to bottom). }
    \label{fig:SAME MI TD XY}
\end{figure*}

The mass distribution of the noBH-QSG-D and wBH-QSG-D streams in the $x-y$ plane is displayed in Fig. \ref{fig:SAME MI TD XY}. We observe that, as in the linear density profile, the mass is concentrated closer to the progenitor's position in the wBH case than in the noBH case. In addition, due to the differing mass-loss rates, we observe that near the centre of the stream, the wBH stream is wider and more offset than the noBH stream because when the stars that composed this section of the stream escaped the progenitor GC the cluster mass was higher. However, due to the weak dependence of the width and radial offset ($\propto M^{1/3}$), this difference is relatively small.

\bsp	
\label{lastpage}
\end{document}